\documentclass[onecollarge]{svjour2}
\smartqed
\usepackage{graphicx,cite,amsmath,amssymb,dsfont}

\newcommand{\be}{\begin{equation}}
\newcommand{\ee}{\end{equation}}
\newcommand{\bea}{\begin{eqnarray}}
\newcommand{\eea}{\end{eqnarray}}
\newcommand{\bd}{\begin{displaymath}}
\newcommand{\ed}{\end{displaymath}}
\newcommand{\vsp}{\vspace*{3mm}}

\newcommand{\bra    }{\langle}
\newcommand{\ket    }{\rangle}

\newcommand{\order}{{\cal O}}

\newcommand{\bc}{\ensuremath{\mathbf{c}}}
\newcommand{\bk}{\ensuremath{\mathbf{k}}}

 \newcommand{\one}{{\rm 1\!\!I}}
 \newcommand{\N}{{\rm I\!N}}

\newcommand{\bcirc}{\circle*{13}}

\newcommand{\here}{\makebox(0,0)}
\newcommand{\notdelta}{\overline{\delta}}

\newcommand{\rme}{{\rm e}}
\newcommand{\rmd}{{\rm d}}
\begin{document}

\title{Constrained Markovian dynamics of random graphs}

%\subtitle{Do you have a subtitle?\\ If so, write it here}

%\titlerunning{Short form of title}        % if too long for running head

\author{A.C.C. Coolen \and A. De Martino \and A. Annibale  %etc.
}

\institute{Ton Coolen, Alessia Annibale  \at
              Department of Mathematics and Randall Division\\
              King's College London\\
              The Strand, London WC2R 2LS, UK\\
              \email{ton.coolen@kcl.ac.uk, alessia.annibale@kcl.ac.uk}           %  \\
           \and
           Andrea De Martino \at
              CNR/INFM (SMC), Dipartimento di Fisica\\
              Sapienza Universit\`a di Roma\\
              p.le A. Moro 2, 00185 Roma, Italy\\
              \email{andrea.demartino@roma1.infn.it}
}

%\date{version of May 25th 2009}
%\date{Received: date / Accepted: date}

\maketitle

\begin{abstract}
We introduce a statistical mechanics formalism for the study of constrained graph evolution as a Markovian stochastic process, in analogy with that available for spin systems, deriving its basic properties and highlighting the role of the `mobility' (the number of allowed moves for any given graph). As an application of the general theory we analyze the properties of degree-preserving Markov chains based on elementary edge switchings. We give an exact yet simple formula for the mobility in terms of the graph's adjacency matrix and its spectrum. This formula allows us to define acceptance probabilities for edge switchings, such that the Markov chains become controlled Glauber-type detailed balance processes, designed to evolve to any required invariant measure (representing the asymptotic frequencies with which the allowed graphs are visited during the process). As a corollary we also derive a condition in terms of simple degree statistics, sufficient to guarantee that, in the limit where the number of nodes diverges,  even for state-independent acceptance probabilities of proposed moves the invariant measure of the process will be uniform. We test our theory on synthetic graphs and on realistic larger graphs as studied in cellular biology.

\keywords{Graph theory \and Stochastic processes \and Edge switching}
%\PACS{PACS code1 \and PACS code2 \and more}
% \subclass{MSC code1 \and MSC code2 \and more}

\end{abstract}

\tableofcontents

%%%%%%%%%%%%%%%%%%%%%%%%%%%%%%%%%%%

\clearpage
\section{Introduction}

The aim of this paper is to develop a mathematical framework for the study of stochastically evolving graphs, defined in terms of generic Markov chains that describe constrained edge re-wiring, and to analyze within this framework the Markovian edge swap dynamics as the simplest non-trivial example of a constrained stochastic graph dynamics.

The formalism which we present can be seen as the analogue of the one available for interacting spin systems. In the latter case, the elementary move is typically a single spin flip, and constrained dynamics have been considered, e.g. by simultaneously flipping pairs of oppositely oriented spins to preserve the total magnetization \cite{kawa,bray}. In extending this theory to evolving graphs we are motivated by the necessity to bridge an existing gap between the static and the dynamical treatment of graphs. While many statistical aspects of random graph topology are now understood (like the influence of topology on processes occurring on graphs, percolation and critical phenomena, loop statistics, or the entropies of different topologies in various random graph ensembles \cite{bara2,bara,new1,doro,new,review,bb1,bbb}), much less work has been invested in the mathematical study of the
dynamics of graphical structures (see \cite{pina,gina,robins} for recent examples). Besides their mathematical interest, dynamical problems are prominent in application areas where the issue of sampling uniformly the space of graphs with certain prescribed macroscopic properties is vital. Multiple examples of these are found in systems biology, where different graph randomization protocols have for instance been tested and used to identify the elementary bricks (`motifs') characterizing such networks, such as the transcriptional regulation network of the bacterium {\it E. coli} \cite{shen-orr,milo,milo2}. Many other examples can be found in economics, ecology and the social sciences, see e.g. \cite{snijders1,snijders2}. At the level where the constraints involve only the simplest quantities, i.e. the degree sequence, even generating such graphs is known to be a non-trivial problem that has produced much inspiring work and some hard open questions \cite{bender,molloy,newman,chung}. It is known, for instance, that a Markov chain based on degree-preserving edge swaps (`switchings') starting from a given graph does not generically produce a uniform sampling of the space of graphs with the same degree sequence, and heuristics with various degrees of sophistication and effectiveness have been employed to restore the uniform measure \cite{rao,Gka,Vig,Sta,chen,Catanzaro,Serrano,Foster,verhelst1}. When more complicated observables  than the degree sequence are involved (like the degree correlations or the number of loops of a given length through each node) the situation rapidly becomes more difficult, both mathematically and numerically.

Here we adopt a top-down approach, and treat the problem of constrained Markovian graph evolution first as generally as possible before dealing with  specific instances. Markov chains describing Glauber-type controlled equilibrium graph dynamics for generic invertible elementary moves are constructed and analyzed in Section 2, including their equilibration properties. Section 3 then focuses on a particular class of moves, namely degree-preserving edge swaps, both in view of their wide application in computer studies and because they are the simplest degree-constrained moves. Among other results, we establish a simple formula for the number of allowed switchings from a given graph configuration (the graph's `mobility') and quantify the dynamical relevance of mobility-related entropic effects. This formula allows us to derive the correct acceptance probabilities for randomly drawn candidate edge swaps that guarantee degree-constrained stochastic evolution towards any desired stationary measure, as well as a condition on degree statistics sufficient to ensure that for uniform acceptances of random edge-swaps the asymptotic measure over the space of allowed graphs generated by the dynamics becomes uniform (implying absence of mobility-related entropic effects). Our results are illustrated and validated via application to various synthetic and biological graphs.

%%%%%%%%%%%%%%%%%%%%%%%%%%%%%%%%%%%%

\section{Formalism and general properties of controlled Markovian graph dynamics}

%%%%%%%%%%%%%%%%%%%%%%%%%%%%%%%%

\subsection{Basic definitions}

We study graphs consisting of $N$ nodes (labeled by Roman indices $i=1\ldots N$)
that can be linked by undirected bonds. A graph is defined microscopically by its adjacency matrix $\bc=\{c_{ij}\}$,  where $c_{ij}=1$ if and only if nodes $i$ and $j$ are connected and $c_{ij}=0$ otherwise, and it is assumed that $c_{ij}=c_{ji}$ and $c_{ii}=0$ for all $(i,j)$. We denote the discrete set of all such graphs as $\mathcal{C}=\{0,1\}^{\frac{1}{2}N(N-1)}$.
Our aim is to define and study constrained Markov chains for the evolution of $\bc$ in some subspace  $\Omega\subseteq\mathcal{C}$, i.e. discrete-time stochastic processes for the probability $p_t(\bc)$ of observing graph $\bc$ at time $t$ of the type
 \begin{eqnarray}
\forall\bc\in\Omega~~~:~~~ p_{t+1}(\bc)&=&\sum_{\bc^\prime\in\Omega} W(\bc|\bc^\prime)p_t(\bc^\prime)
 \label{eq:MarkovChain}
 \end{eqnarray}
 with $t\in\N$, $W(\bc|\bc^\prime)\geq 0$ $\forall \bc,\bc^\prime\in\Omega$, and $\sum_{\bc\in\Omega}W(\bc|\bc^\prime)=1$ $\forall\bc^\prime\in\Omega$.  Here the quantity $W(\bc|\bc^\prime)$ denotes the single-step transition probability from graph $\bc^\prime$ to graph $\bc$.
We focus on processes of the form (\ref{eq:MarkovChain}) that have the following additional properties:

 \begin{enumerate}

  \item[~(i)] The process (\ref{eq:MarkovChain}) allows only for a given limited set $\Phi$ of elementary moves $F:\Omega_F\to \Omega$, which  are constrained, in that each $F$ can act only on a subset $\Omega_F\subseteq \Omega$ of all possible graphs. \vsp

  \item[~(ii)] The process (\ref{eq:MarkovChain}) converges to the invariant measure $p_\infty(\bc)=Z^{-1}\rme^{-H(\bc)}$ on $\Omega$, in which $H$ is a prescribed function,  for any choice of initial conditions $p_0(\bc)$. Here $Z=\sum_{\bc\in\Omega}\rme^{-H(\bc)}$ to ensure normalization.

  \end{enumerate}

\noindent
 For each elementary move $F\in\Phi$ we define an indicator
function $I_F(\bc)\in\{0,1\}$, where $I_F(\bc)=1$ if and only if the move $\bc\to F\bc$
is allowed. In addition to the above requirements we will also demand that all elementary moves are invertible, i.e.

\begin{enumerate}
\item[~(iii)] For each $F\in\Phi$ there exists a unique $F^{-1}\in\Phi$ such that $FF^{-1}=F^{-1}F=\one$.
Both $F$ and $F^{-1}$ are taken to act on the same subset of states, i.e. $I_F(\bc)=I_{F^{-1}}(\bc)$ for all $\bc\in\Omega$.
\end{enumerate}

\noindent
Processes of the form (\ref{eq:MarkovChain}) can be used to generate graphs with controlled properties, such as built-in constraints (which can be encoded in the subset $\Omega$ of allowed graphs, and induce the limitations on the applicability of moves that generate the subsets $\Omega_F$) and specific statistical weights (which can be encoded in the invariant measure $p_\infty(\bc)$, i.e. in the function $H(\bc)$). The scenario of having a limited set of elementary moves, which each can act only on certain configurations $\bc$, describes many of the commonly studied dynamical processes for graphs. In the examples that will be worked out explicitly in subsequent sections, $\Phi$ is the set of elementary moves that preserve the degree $k_i(\bc)=\sum_j c_{ij}$ of every node, the simplest possible such move $F$ being an edge-swap between two pairs of nodes.

Our next task is, for any given subset $\Omega\subseteq\mathcal{C}$ of states, any given set $\Phi$ of possible moves with state-dependent application constraints that meet conditions (i,iii), and any given measure $p_\infty(\bc)$ of the form (ii), to construct appropriate transition probabilities $W(\bc|\bc^\prime)$ such that the state probabilities $p_t(\bc)$ generated by the Markov chain  (\ref{eq:MarkovChain}) are guaranteed to converge to the desired values
$p_\infty(\bc)$.

%%%%%%%%%%%%%%%%%%%%%%%%%

\subsection{The Markov chain transition probabilities}

In order to construct the transition probabilities $W(\bc|\bc^\prime)$, we can resort to the familiar ideas behind Monte-Carlo (or Glauber-type) processes for the simulation of physical systems, provided these are properly adapted to build in the constraints on the applicability of the allowed graph transitions $F\in \Phi$. In particular, we will construct our transition probabilities such that the corresponding process obeys detailed balance, i.e.
\begin{eqnarray}
\forall\bc,\bc^\prime\in\Omega~~~:~~~W(\bc|\bc^\prime)p_\infty(\bc^\prime)=W(\bc^\prime|\bc)p_\infty(\bc)
\label{eq:DetailedBalance}
\end{eqnarray}
Summation over $\bc^\prime\in\Omega$ in (\ref{eq:DetailedBalance}) reveals in the usual manner that
(\ref{eq:DetailedBalance}) implies stationarity of $p_\infty$, i.e. $\forall \bc\in\Omega:$ $\sum_{\bc^\prime\in\Omega}W(\bc|\bc^\prime)p_\infty(\bc^\prime)=p_\infty(\bc)$.
 We next define a new set $\Phi^\prime$ of moves which excludes the identity operation:
 \begin{equation}
 \Phi^\prime=\{F\in\Phi|~\exists \bc\in\Omega~{\rm such~that}~F\bc\neq \bc\}
 \end{equation}
A generic detailed balance Markov chain is then obtained by choosing
 \begin{eqnarray}
W(\bc|\bc^\prime)&=& \sum_{F\in\Phi^\prime} q(F|\bc^\prime)\Big[\delta_{\bc,F\bc^\prime} A(F\bc^\prime|\bc^\prime)+\delta_{\bc,\bc^\prime} [1-A(F\bc^\prime|\bc^\prime)]\Big]
\label{eq:TransitionProbabilities}
\end{eqnarray}
The rationale and interpretation of this choice (\ref{eq:TransitionProbabilities}) is as follows. At each step a candidate move $F\in\Phi^\prime$ is drawn with some probability $q(F|\bc^\prime)$, where $\bc^\prime$ denotes the current state. This move is accepted (and the transition $\bc^\prime\to\bc=F\bc^\prime$ is executed) with some probability $A(F\bc^\prime|\bc^\prime)\in[0,1]$, which depends on both the current state $\bc^\prime$ and on the proposed new state $F\bc^\prime$. If the move is rejected, which happens with probability $1\!-\!A(F\bc^\prime|\bc^\prime)$, the system remains in the current state $\bc^\prime$.
Clearly (\ref{eq:TransitionProbabilities}) obeys $\sum_{\bc\in\Omega} W(\bc|\bc^\prime)=1$ for all $\bc^\prime\in\Omega$, as it should.
Working out the detailed balance condition (\ref{eq:DetailedBalance}), upon writing the equilibrium state in the Boltzmann form $p_\infty(\bc)=Z^{-1}\exp[-H(\bc)]$, leads to the following conditions for $q(F|\bc)$ and $A(\bc|\bc^\prime)$:
\begin{eqnarray}
(\forall \bc\!\in\!\Omega)(\forall F\!\in\!\Phi^\prime):&~~~&
q(F|\bc) A(F\bc|\bc)
\rme^{-H(\bc)}
= q(F^{-1}|F\bc) A(\bc|F\bc)
\rme^{-H(F\bc)}
\end{eqnarray}
Now let $n(\bc)$ denote the number of moves that can act on a state $\bc$ (to which we shall refer as the `mobility' of state $\bc$), defined as
\begin{equation}
n(\bc)=\sum_{F\in\Phi^\prime} I_F(\bc).
\end{equation}
The fact that for constrained moves these numbers $n(\bc)$ are generally state-dependent, forces us to choose Monte-Carlo acceptance
probabilities $A(\bc|\bc^\prime)$ that no longer depend on $H(\bc)-H(\bc^\prime)$ only.
If the candidate moves $F$ are drawn randomly and with equal probabilities from those that are allowed
to act, we have  $q(F|\bc)=I_F(\bc)/n(\bc)$ and the detailed balance condition becomes
\begin{eqnarray}
(\forall \bc\!\in\!\Omega)(\forall F\!\in\!\Phi^\prime):&~~~&
 A(F\bc|\bc)
\rme^{-H(\bc)}/n(\bc)= A(\bc|F\bc)
\rme^{-H(F\bc)}/ n(F\bc)
\end{eqnarray}
It is then clear that having a state-dependent $n(\bc)$ is equivalent to modifying the state energies, viz.
$H(\bc)\to H(\bc)+\log n(\bc)$, so that the Monte-Carlo acceptance rates can be chosen as
\begin{eqnarray}
A(\bc|\bc^\prime)=
\frac{n(\bc^\prime)\rme^{-\frac{1}{2}[H(\bc)-H(\bc^\prime)]}}{
n(\bc^\prime)e^{-\frac{1}{2}[H(\bc)-H(\bc^\prime)]}+
n(\bc)\rme^{\frac{1}{2}[H(\bc)-H(\bc^\prime)]}}
\label{eq:FinalAcceptanceRates}
\end{eqnarray}
(note that move reversibility implies $n(\bc)\geq 1$) and the end result is the Markov chain defined by the transition probabilities
\begin{eqnarray}
\hspace*{-0mm}
W(\bc|\bc^\prime)&=& \sum_{F\in\Phi^\prime} \frac{I_F(\bc^\prime)}{n(\bc^\prime)}
\Big[
\frac{\delta_{\bc,F\bc^\prime}  n(\bc^\prime)\rme^{-\frac{1}{2}[H(F\bc^\prime)-H(\bc^\prime)]}
+\delta_{\bc,\bc^\prime} n(F\bc^\prime)\rme^{\frac{1}{2}[H(F\bc^\prime)-H(\bc^\prime)]}}{
n(\bc^\prime)\rme^{-\frac{1}{2}[H(F\bc^\prime)-H(\bc^\prime)]}+
n(F\bc^\prime)\rme^{\frac{1}{2}[H(F\bc^\prime)-H(\bc^\prime)]}}
\Big]
\label{eq:FinalTransitionProbabilities}
\end{eqnarray}
One easily confirms by direct substitution that the transition probabilities (\ref{eq:FinalTransitionProbabilities}) indeed
define a Markov process which
leaves the measure $p_\infty(\bc)=Z^{-1}\exp[-H(\bc)]$ invariant,
since for any $\bc\in\Omega$ we find, using simple identities such as $I_F(\bc)=I_{F^{-1}}(\bc)=I_F(F\bc)$ and $\sum_{F\in\Phi^\prime}G(F)=\sum_{F\in\Phi^\prime}G(F^{-1})$, that
\begin{eqnarray}
\sum_{\bc^\prime\in\Omega}W(\bc|\bc^\prime)p_\infty(\bc^\prime)&=&\frac{1}{Z}\sum_{\bc^\prime\in\Omega}W(\bc|\bc^\prime)\rme^{-H(\bc^\prime)}
\nonumber
\\[-2mm]
\hspace*{-23mm}&&
\hspace*{-15mm}=\frac{\rme^{-H(\bc)}}{Z}\sum_{F\in\Phi^\prime}  \sum_{\bc^\prime\in\Omega} \frac{I_F(\bc^\prime)}{n(\bc^\prime)}
\Big[
\frac{\delta_{F^{-1}\bc,\bc^\prime}  n(\bc^\prime)\rme^{-\frac{1}{2}[H(\bc^\prime)-H(F\bc^\prime)]}
+\delta_{\bc,\bc^\prime} n(F\bc^\prime)\rme^{\frac{1}{2}[H(F\bc^\prime)-H(\bc^\prime)]}}{
n(\bc^\prime)\rme^{-\frac{1}{2}[H(F\bc^\prime)-H(\bc^\prime)]}+
n(F\bc^\prime)\rme^{\frac{1}{2}[H(F\bc^\prime)-H(\bc^\prime)]}}
\Big]
\nonumber
\\
\hspace*{-23mm}
&=&p_\infty(\bc)\sum_{F\in\Phi^\prime} \frac{I_F(\bc)}{n(\bc)}\left\{
\frac{  n(\bc)\rme^{-\frac{1}{2}[H(F^{-1}\bc)-H(\bc)]}
}{
n(F^{-1}\bc)\rme^{-\frac{1}{2}[H(\bc)-H(F^{-1}\bc)]}+
n(\bc)\rme^{\frac{1}{2}[H(\bc)-H(F^{-1}\bc)]}}
\right.\nonumber
\\
\hspace*{-23mm}
&&
\left.
\hspace*{39mm}
+
\frac{
 n(F\bc)\rme^{\frac{1}{2}[H(F\bc)-H(\bc)]}}{
n(\bc)\rme^{-\frac{1}{2}[H(F\bc)-H(\bc)]}+
n(F\bc)\rme^{\frac{1}{2}[H(F\bc)-H(\bc)]}}
\right\}
\nonumber
\\
\hspace*{-23mm}
&=&
p_\infty(\bc)\sum_{F\in\Phi^\prime} \frac{I_F(\bc)}{n(\bc)}~=~p_\infty(\bc)
\end{eqnarray}

%%%%%%%%%%%%%%%%%%%%%%%%%%%

\subsection{Master equation representation of the process}

The process defined by (\ref{eq:MarkovChain},\ref{eq:FinalTransitionProbabilities}) allows for relatively easy numerical implementation, but for mathematical analysis a real-time formulation in the form of a master equation is more convenient. The formal method to go from a process of the
form (\ref{eq:MarkovChain}) to a master equation, is to assume that the {\em duration} of each of the discrete iteration steps in (\ref{eq:MarkovChain}) is a continuous random number \cite{vk}. The statistics of these
random durations are defined by  the probability $\pi_m(t)$ that at time $t\geq 0$ precisely $m$ iteration steps have been made. Our new real-time process is now described by
\begin{eqnarray}
P_t(\bc)&=&\sum_{m\geq 0}\pi_m(t)p_m(\bc)
=
\sum_{m\geq 0}\pi_m(t)
\sum_{\bc^{\prime}\in\Omega}(W^m)(\bc|\bc^{\prime})p_0(\bc^{\prime})
\end{eqnarray}
where the time $t$ has now become a continuous variable. For $\pi_m(t)$ we make the
Poissonnian choice
$\pi_m(t)=(t/\tau)^m \rme^{-t/\tau}/m!$,
with the properties
\be
\frac{\rmd}{\rmd t}\pi_{m>0}(t)=\tau^{-1}\big[\pi_{m-1}(t)-\pi_m(t)\big],~~~~~~
\frac{\rmd}{\rmd t}\pi_{0}(t)=-\tau^{-1}\pi_m(t)
\ee
From $\bra m\ket_\pi=t/\tau$ it follows that $\tau$ is the average
duration of a single discrete iteration step.
Our choice for $\pi_m(t)$ allows us to write for the time
derivative of $P_t(\bc)$:
\begin{eqnarray}
\hspace*{-0mm}
\tau\frac{\rmd}{\rmd t}P_t(\bc)&=&\sum_{m>0}\pi_{m-1}(t)
\sum_{\bc^{\prime}\in\Omega}(W^m)(\bc|\bc^{\prime})p_0(\bc^{\prime})
-\!\sum_{m\geq 0}\pi_m(t)
\sum_{\bc^{\prime}\in\Omega}(W^m)(\bc|\bc^{\prime})p_0(\bc^{\prime})
\\
\hspace*{-0mm}
&=&- P_t(\bc)+
\sum_{\bc^{\prime}\in\Omega}W(\bc|\bc^{\prime})P_t(\bc^{\prime})
\label{eq:MasterEqn}
\end{eqnarray}
which has the form of a master equation.
Applying this procedure to our present Markov chain (\ref{eq:MarkovChain},\ref{eq:FinalTransitionProbabilities}) gives its corresponding master equation which (after re-arranging of terms) takes the transparent form
\begin{eqnarray}
\tau\frac{\rmd}{\rmd t}P_t(\bc)&=&
\sum_{F\in\Phi^\prime}I_F(\bc)
\left\{
\frac{1}{n(F\bc)}
\frac{ n(F\bc)\rme^{\frac{1}{2}[H(F\bc)-H(\bc)]}
P_t(F\bc)
}{
n(\bc)\rme^{-\frac{1}{2}[H(F\bc)-H(\bc)]}+
n(F\bc)\rme^{\frac{1}{2}[H(F\bc)-H(\bc)]}}
\nonumber
\right.
\\
&&\left.\hspace*{30mm}-\frac{1}{n(\bc)}\frac{n(\bc)\rme^{-\frac{1}{2}[H(F\bc)-H(\bc)]}P_t(\bc)
}{
n(\bc)\rme^{-\frac{1}{2}[H(F\bc)-H(\bc)]}+
n(F\bc)\rme^{\frac{1}{2}[H(F\bc)-H(\bc)]}}
\right\}
\nonumber
\\
&=&
\sum_{F\in\Phi^\prime}I_F(\bc)
\left\{
\frac{w^+_{F}(\bc)}{n(F\bc)}P_t(F\bc)
-\frac{w^-_F(\bc)}{n(\bc)}P_t(\bc)\right\}
\label{eq:general_master}
\end{eqnarray}
where, using the the short-hand $\Delta_F U(\bc)=U(F\bc)-U(\bc)$,  we have identified the transition rates
\begin{eqnarray}
w^\pm_F(\bc)&=& \frac{1}{2}\pm\frac{1}{2}\tanh\Big[\frac{1}{2}\Delta_F[H(\bc)+\log n(\bc)]\Big]
\label{eq:transition_rates}
\end{eqnarray}
For $N\to\infty$ there can be no difference between the process (\ref{eq:general_master}), describing
random durations of the steps of
 the Markov chain, and one where the duration of {\em each} step would be $\tau$ (rather than just their average). This follows from
 the moments of the Poisson process, viz. $\bra m^2\ket_\pi=\bra m\ket^2_\pi+\bra m\ket_\pi$, which
 guarantee that for finite real time $t$ the relative uncertainty in the number of iterations will vanish
 according to $\sqrt{\bra m^2\ket_\pi-\bra m\ket_\pi^2}/\bra m\ket_\pi=\order(N^{-1/2})$.
 If we were to repeat the above derivation for a process where {\em each} randomly chosen and possible edge swap
was always accepted, we would again find equation (\ref{eq:general_master}), but now with the trivial
transition rates $w^\pm_F(\bc)=1$ for all $\bc\in\Omega$ and all $F\in\Phi^\prime$.

From (\ref{eq:general_master},\ref{eq:transition_rates}) one can proceed to derive convenient dynamical equations for ensemble averages.
These are particularly compact and transparent for those situations where $F^{-1}=F$ for all $F\in \Phi^\prime$, which will, for instance, be true for edge swap graph dynamics. To be specific,
upon defining $\bra f(\bc)\ket=\sum_{\bc\in\Omega}P_t(\bc)f(\bc)$ for any arbitrary observable $f(\bc)$, we obtain, using identities such as $I_F(F\bc)=I_F(\bc)$, $\Delta_F U(F\bc)=-\Delta_F U(\bc)$ and $w^+_F(F\bc)=w^-_F(\bc)$ (which all follow directly from $F^{-1}=F$):
\begin{eqnarray}
\tau\frac{\rmd}{\rmd t}\bra f(\bc)\ket&=& \sum_{F\in\Phi^\prime}\Big\bra \frac{I_F(\bc)}{n(\bc)}
\left\{f(F\bc)w^+_{F}(F\bc)
-f(\bc)w^-_F(\bc)\right\}\Big\ket
\nonumber
\\
&=& \sum_{F\in\Phi^\prime}\Big\bra \frac{I_F(\bc)}{n(\bc)}~w^-_{F}(\bc)~ \Delta_F f(\bc)\Big\ket
\end{eqnarray}

%%%%%%%%%%%%%%%%%%%%%%%

\subsection{Convergence towards equilibrium}

Convergence of (\ref{eq:general_master},\ref{eq:transition_rates}) towards the equilibrium state generally implies that the same is true for the underlying Markov chain (\ref{eq:MarkovChain},\ref{eq:FinalTransitionProbabilities})\footnote{Exceptions to this would be e.g. periodic Markov chains.}. The physically most intuitive convergence proofs are based on constructing a Lyapunov function $F(t)$. Here we choose for $F(t)$ the Kullback-Leibler divergence between the equilibrium state $p_\infty(\bc)=Z^{-1}\exp[-H(\bc)]$ and the instantaneous distribution $P_t(\bc)$, which obeys $F(t)\geq 0$ for all $t$ by definition, and is zero only if the measures $P_{t}(\bc)$ and $p_{\infty}(\bc)$  are identical:
\begin{eqnarray}
F(t)&=& \sum_{\bc\in\Omega}P_t(\bc)\log [P_t(\bc)/p_\infty(\bc)]=\log Z+\sum_{\bc\in\Omega}P_t(\bc)[H(\bc)+\log P_t(\bc)]
\end{eqnarray}
The proof that $F(t)$ decreases monotonically is standard and relies only on the detailed balance condition (\ref{eq:DetailedBalance}) and on the normalization $\sum_{\bc\in\Omega}W(\bc|\bc^\prime)=1$ $\forall\bc^\prime\in\Omega$. We repeat it here only for completeness, and to show that it can handle the constraints in the graph dynamics in a straightforward manner:
\begin{eqnarray}
\hspace*{-0mm}
\frac{\rmd}{\rmd t}F(t)&=& \sum_{\bc\in\Omega}[H(\bc)+\log P_t(\bc)]\frac{\rmd}{\rmd t}P_t(\bc)
\nonumber\\
&=&
\hspace*{-0mm}
\frac{1}{\tau}\sum_{\bc\in\Omega}[H(\bc)+\log P_t(\bc)]\Big[\sum_{\bc^\prime\in\Omega}W(\bc|\bc^\prime)P_t(\bc^\prime)-P_t(\bc)\Big]
\nonumber\\
&=&
\hspace*{-0mm}
\frac{1}{\tau}\sum_{\bc\in\Omega}[H(\bc)+\log P_t(\bc)]\sum_{\bc^\prime\in\Omega}\Big[W(\bc|\bc^\prime)P_t(\bc^\prime)-W(\bc^\prime|\bc)P_t(\bc)\Big]
\nonumber\\
&=&
\hspace*{-0mm}
\frac{1}{2\tau}\sum_{\bc,\bc^\prime\in\Omega}\Big[[H(\bc)\!+\!\log P_t(\bc)]-[H(\bc^\prime)\!+\!\log P_t(\bc^\prime)]\Big]
\Big[W(\bc|\bc^\prime)P_t(\bc^\prime)-W(\bc^\prime|\bc)P_t(\bc)\Big]
\nonumber\\
&=&
\hspace*{-0mm}
\frac{1}{2\tau}\sum_{\bc,\bc^\prime\in\Omega}\Big[[H(\bc)\!+\!\log P_t(\bc)]-[H(\bc^\prime)\!+\!\log P_t(\bc^\prime)]\Big]\nonumber
\\
&&\hspace*{20mm} \times \Big[ \rme^{H(\bc^\prime)+\log P_t(\bc^\prime)}W(\bc|\bc^\prime)\rme^{-H(\bc^\prime)}-\rme^{H(\bc)+\log P_t(\bc)}
W(\bc^\prime|\bc)\rme^{-H(\bc)}\Big]
\end{eqnarray}
We next use the detailed balance identity (\ref{eq:DetailedBalance}) and obtain
\begin{eqnarray}
\hspace*{-0mm}
\frac{\rmd}{\rmd t}F(t)&=&
-\frac{1}{2\tau}\sum_{\bc,\bc^\prime\in\Omega}W(\bc|\bc^\prime)\rme^{-H(\bc^\prime)}
\Big[[H(\bc)\!+\!\log P_t(\bc)]-[H(\bc^\prime)\!+\!\log P_t(\bc^\prime)]\Big]\nonumber
\\
&&\hspace*{20mm}\times \Big[\rme^{H(\bc)+\log P_t(\bc)} -\rme^{H(\bc^\prime)+\log P_t(\bc^\prime)}\Big]~\leq~0
\label{eq:Hdecreases}
\end{eqnarray}
The last step derives from the general identity $(e^x-e^y)(x-y)\geq 0$ for all $(x,y)$, with equality only if $x=y$.
Since $F(t)$ is bounded from below, it is indeed a Lyapunov function for the process (\ref{eq:general_master}).
The distance between $P_t(\bc)$ and the equilibrium measure $p_\infty(\bc)=Z^{-1}e^{-H(\bc)}$ decreases monotonically until $\rmd F(t)/\rmd t=0$.
Inspection of the conditions for equality in (\ref{eq:Hdecreases}) shows that $F(t)$ stops  decreasing only if $P_t(\bc)$ has reached a point $P(\bc)=\chi(\bc)\rme^{-H(\bc)}$ with:
 \begin{eqnarray}
 (\forall \bc,\bc^\prime\in\Omega):&~~& W(\bc|\bc^\prime)=0~~~{\rm or}~~~\chi(\bc)=\chi(\bc^\prime)
\label{eq:fixed_point}
 \end{eqnarray}
 If we define $\Omega_{\bc}=\{\bc^\prime\!\in\Omega |~\exists \ell\in\N:~(W^\ell)(\bc^\prime|\bc)>0\}$ (representing
the set of  all states that are accessible from $\bc$ via repeated iteration of the Markov chain), then for all $\bc^\prime\in \Omega_\bc$ we must have $\chi(\bc^\prime)=\chi(\bc)$.
 Each solution of (\ref{eq:fixed_point}) generates an invariant measure under the dynamics, since substitution of $P(\bc)=\chi(\bc)\rme^{-H(\bc)}$
 in both sides of (\ref{eq:MasterEqn}) shows, using (\ref{eq:DetailedBalance}),
 \begin{eqnarray}
 \sum_{\bc^\prime\in\Omega}W(\bc|\bc^\prime)\chi(\bc^\prime)\rme^{-H(\bc^\prime)}-\chi(\bc)\rme^{-H(\bc)}
 &=& \rme^{-H(\bc)}\Big[\sum_{\bc^\prime\in\Omega}\chi(\bc^\prime) W(\bc^\prime|\bc)\!-\!\chi(\bc)\Big]\nonumber
 \\
 &=&
 \rme^{-H(\bc)}\chi(\bc)\Big[\sum_{\bc^\prime\in\Omega} W(\bc^\prime|\bc)\!-\!1\Big]=0
\end{eqnarray}
 One such stationary solution $P(\bc)$ is the equilibrium measure, corresponding to $\chi(\bc)=Z^{-1}$ for all $\bc\in\Omega$.
 If our Markov chain is ergodic, i.e. if $\Omega_\bc=\Omega$ for all $\bc\in\Omega$, then it is the {\em only} such state to satisfy (\ref{eq:fixed_point}), and our process must always evolve towards $p_\infty(\bc)=Z^{-1}\exp[-H(\bc)]$.

We shall next focus our attention on a specific class of moves that preserve the values of all node degrees $k_i(\bc)=\sum_j c_{ij}$, and work out the details of the corresponding equilibrium dynamics.

%%%%%%%%%%%%%%%%%%%%

\section{Degree-constrained dynamics}

%%%%%%%%%%%%%%%%%%%%%%%%%

\subsection{Elementary moves defined on the basis of `edge swaps'}

Here we use the results of the previous section to construct stochastic processes for evolving graphs which preserve their full degree sequence $\bk(\bc)=\{k_i(\bc)\}$, and in addition evolve (within the subspace
of graphs with fixed degree sequence $\bk$) to any desired
prescribed measure of the form $p_\infty(\bc)=Z^{-1}\exp[-H(\bc)]$.
Let $\Omega[\bk]$ denote  the discrete set of all un-directed graphs that have the specified degree sequence $\bk$:
\begin{eqnarray}
\Omega[\bk]=\{\bc\in\{0,1\}^{\frac{1}{2}N(N-1)}|~k_i(\bc)=k_i~~\forall i=1\ldots N\}
\end{eqnarray}
This set will play the role of our state space $\Omega$ in (\ref{eq:MarkovChain}).
For the elementary moves of our Markov chain we choose the so-called `edge swaps' (simple cases of the more general family of Seidel switches \cite{seidel}), which are the simplest
possible graph transitions that conserve the values of all degrees (see Appendix \ref{app:switches}). An edge swap is defined by the following protocol:
\begin{description}
\item~(a)~ draw four distinct nodes $(i,j,k,\ell)$
\item~(b)~ check whether $(c_{i\ell},c_{jk},c_{ij},c_{k\ell})=(1,1,0,0)$
\item~(c)~ if yes, invert these four variables: $(c_{i\ell},c_{jk},c_{ij},c_{k\ell})\to (0,0,1,1)$
\end{description}
We generate the set $\Phi^\prime$ of such moves by summing uniquely over all quadruplets of nodes, with validity checks for all possible edge
swaps that could be carried out for each quadruple. To do this carefully
we define $Q=\{(i,j,k,\ell)\in\{1,\ldots,N\}^4|~i\!<\!j\!<\!k\!<\!\ell\}$, and we
no longer allow for any permutations of the four nodes.
The six potential edge swaps are then found to be the following, with thick lines indicating existing links, and thin lines indicating absent links that will be swapped with the existing ones:
\vspace*{12mm}

\setlength{\unitlength}{0.140mm}\hspace*{-9mm}

 \begin{picture}(500,100)(-250,0)
 \put(50,175){\here{I}}
 \put(5,0){\here{\bcirc}}\put(105,0){\here{\bcirc}}\put(5,100){\here{\bcirc}}\put(105,100){\here{\bcirc}}
 \put(0,125){\here{$i$}}\put(100,125){\here{$j$}}\put(100,-25){\here{$k$}}\put(0,-25){\here{$\ell$}}
 \thinlines \put(0,0){\line(0,1){100}} \put(100,0){\line(0,1){100}}
 \thicklines\put(0,100){\line(1,0){100}}\put(0,101){\line(1,0){100}}\put(0,99){\line(1,0){100}}
 \put(0,0){\line(1,0){100}} \put(0,-1){\line(1,0){100}}\put(0,1){\line(1,0){100}}

 \put(250,175){\here{II}}
  \put(205,0){\here{\bcirc}}\put(305,0){\here{\bcirc}}\put(205,100){\here{\bcirc}}\put(305,100){\here{\bcirc}}
 \put(200,125){\here{$i$}}\put(300,125){\here{$j$}}\put(300,-25){\here{$k$}}\put(200,-25){\here{$\ell$}}
 \thinlines \put(200,0){\line(1,1){100}} \put(300,0){\line(-1,1){100}}
 \thicklines \put(200,100){\line(1,0){100}} \put(200,101){\line(1,0){100}} \put(200,99){\line(1,0){100}}
 \put(200,0){\line(1,0){100}}\put(200,1){\line(1,0){100}}\put(200,-1){\line(1,0){100}}

 \put(450,175){\here{III}}
  \put(405,0){\here{\bcirc}}\put(505,0){\here{\bcirc}}\put(405,100){\here{\bcirc}}\put(505,100){\here{\bcirc}}
 \put(400,125){\here{$i$}}\put(500,125){\here{$j$}}\put(500,-25){\here{$k$}}\put(400,-25){\here{$\ell$}}
 \thinlines \put(400,0){\line(0,1){100}} \put(500,0){\line(0,1){100}}
 \thicklines \put(400,100){\line(1,-1){100}}\put(400,101.5){\line(1,-1){100}}\put(400,98.5){\line(1,-1){100}}
 \put(400,0){\line(1,1){100}}\put(400,1.5){\line(1,1){100}}\put(400,-1.5){\line(1,1){100}}

\end{picture}
\vspace*{10mm}

\noindent
together with their inverse edge swaps

\vspace*{12mm}

\setlength{\unitlength}{0.140mm}\hspace*{-9mm}

  \begin{picture}(500,100)(-250,0)
 \put(50,175){\here{IV}}
 \put(5,0){\here{\bcirc}}\put(105,0){\here{\bcirc}}\put(5,100){\here{\bcirc}}\put(105,100){\here{\bcirc}}
 \put(0,125){\here{$i$}}\put(100,125){\here{$j$}}\put(100,-25){\here{$k$}}\put(0,-25){\here{$\ell$}}
 \thicklines \put(0,0){\line(0,1){100}} \put(100,0){\line(0,1){100}}
 \put(1,0){\line(0,1){100}} \put(101,0){\line(0,1){100}}
\put(-1,1){\line(0,1){100}} \put(99,0){\line(0,1){100}}
 \thinlines\put(0,100){\line(1,0){100}} \put(0,0){\line(1,0){100}}

 \put(250,175){\here{V}}
  \put(205,0){\here{\bcirc}}\put(305,0){\here{\bcirc}}\put(205,100){\here{\bcirc}}\put(305,100){\here{\bcirc}}
 \put(200,125){\here{$i$}}\put(300,125){\here{$j$}}\put(300,-25){\here{$k$}}\put(200,-25){\here{$\ell$}}
 \thicklines \put(200,0){\line(1,1){100}} \put(300,0){\line(-1,1){100}}
 \put(201,-1){\line(1,1){100}} \put(299,-1){\line(-1,1){100}}
 \put(199,1){\line(1,1){100}} \put(301,1){\line(-1,1){100}}
 \thinlines \put(200,100){\line(1,0){100}}
 \put(200,0){\line(1,0){100}}

 \put(450,175){\here{VI}}
  \put(405,0){\here{\bcirc}}\put(505,0){\here{\bcirc}}\put(405,100){\here{\bcirc}}\put(505,100){\here{\bcirc}}
 \put(400,125){\here{$i$}}\put(500,125){\here{$j$}}\put(500,-25){\here{$k$}}\put(400,-25){\here{$\ell$}}
 \thicklines \put(400,0){\line(0,1){100}} \put(500,0){\line(0,1){100}}
 \put(401,0){\line(0,1){100}} \put(501,0){\line(0,1){100}}
 \put(399,0){\line(0,1){100}} \put(499,0){\line(0,1){100}}
 \thinlines \put(400,100){\line(1,-1){100}}
 \put(400,0){\line(1,1){100}}

\end{picture}

\vspace*{10mm}

\noindent
This leads to a natural grouping of edge swaps into the three pairs (I,IV), (II,V), and (III,VI). We label all three resulting auto-invertible operations for each ordered quadruple $(i,j,k,\ell)$
by adding a subscript $\alpha$, so that our set $\Phi^\prime$ of
all auto-invertible edge swaps are from now on written as $F_{ijk\ell;\alpha}$ with  $i<j<k<\ell$ and $\alpha\in\{1,2,3\}$. We define suitable associated indicator
functions $I_{ijk\ell;\alpha}(\bc)\in\{0,1\}$ that detect whether (1) or not (0) the edge swap $F_{ijk\ell;\alpha}$ can act on state $\bc$, so
\begin{eqnarray}
I_{ijk\ell;1}(\bc)&=& c_{ij}c_{k\ell}(1-c_{i\ell})(1-c_{jk})+(1-c_{ij})(1-c_{k\ell})c_{i\ell}c_{jk}
\label{eq:Fcond1}
\\
I_{ijk\ell;2}(\bc)&=& c_{ij}c_{k\ell}(1-c_{ik})(1-c_{j\ell})+(1-c_{ij})(1-c_{k\ell})c_{ik}c_{j\ell}
\label{eq:Fcond2}
\\
I_{ijk\ell;3}(\bc)&=& c_{ik}c_{j\ell}(1-c_{i\ell})(1-c_{jk})+(1-c_{ik})(1-c_{j\ell})c_{i\ell}c_{jk}
\label{eq:Fcond3}
\end{eqnarray}
If $F_{ijk\ell;\alpha}$ can indeed act, i.e. if $I_{ijk\ell;\alpha}(\bc)=1$, the edge swap will operate as follows:
\begin{eqnarray}
F_{ijk\ell;\alpha}(\bc)_{qr}&= 1-c_{qr}~~~& {\rm for}~(q,r)\in {\cal S}_{ijk\ell;\alpha}\\
F_{ijk\ell;\alpha}(\bc)_{qr}&= c_{qr}~~~& {\rm for}~(q,r)\notin {\cal S}_{ijk\ell;\alpha}
\end{eqnarray}
\vspace*{-3mm}
where
\vspace*{-3mm}
\begin{eqnarray}
{\cal S}_{ijk\ell;1}&=& \{(i,j),(k,\ell),(i,\ell),(j,k)\}
\label{eq:S1}
\\
{\cal S}_{ijk\ell;2}&=& \{(i,j),(k,\ell),(i,k),(j,\ell)\}
\label{eq:S2}
\\
{\cal S}_{ijk\ell;3}&=& \{(i,k),(j,\ell),(i,\ell),(j,k)\}
\label{eq:S3}
\end{eqnarray}
Each edge swap is its own inverse, so the property $I_F(\bc)=I_{F^{-1}}(\bc)$, that we relied upon several times, is trivially valid.
 The link with the theory in the previous section thus becomes, for arbitrary $G(F)$,
\begin{eqnarray}
\hspace*{-5mm}
F(\bc)&~\to~~& F_{ijk\ell;\alpha}(\bc)\\
\hspace*{-5mm}
I_F(\bc)&~\to~~& I_{ijk\ell;\alpha}(\bc)\\
\hspace*{-5mm}
\Omega_{F}(\bc)&~\to~~& \Omega_{ijk\ell;\alpha}(\bc)=\{\bc\!\in\!\{0,1\}^{\frac{1}{2}N(N-1)}|~I_{ijk\ell;\alpha}(\bc)=1\}\\
\hspace*{-5mm}
\sum_{F\in\Phi^\prime}G(F)&~\to~~& \sum_{i<j<k<\ell}~\sum_{\alpha\leq 3} G(F_{ijk\ell;\alpha})
 \end{eqnarray}

%%%%%%%%%%%%%%%%%%%

\subsection{The number of possible edge swaps - graph mobility}

Given the above definitions,
the number $n(\bc)$ of possible edge swaps that can act on a given graph $\bc$ (its mobility) is given by the following expression
\begin{equation}
n(\bc)=\sum_{\alpha=1}^3\sum_{i<j<k<\ell}I_{ijk\ell;\alpha}(\bc)
\label{eq:nc_starting_point}
\end{equation}
However, it is possible to obtain a simplified formula for $n(\bc)$ in terms mainly of the degree and loop statistics of $\bc$, by exploiting the fact that there are $4!=24$ possible orderings of the four indices $(i,j,k,\ell)$,
which implies that for any fully permutation invariant quantity $\Gamma_{ijk\ell}$ one may always write (with the short-hand $\notdelta_{ij}=1-\delta_{ij}$):
\begin{equation}
\sum_{i<j<k<\ell}\Gamma_{ijk\ell}=
\frac{1}{24}\sum_{ijk\ell}\notdelta_{ij}\notdelta_{ik}\notdelta_{i\ell}\notdelta_{jk}\notdelta_{j\ell}\notdelta_{k\ell}\Gamma_{ijk\ell}
\end{equation}
 We can use this to write (\ref{eq:nc_starting_point}), using $c_{ii}=0$ for all $i$, in the alternative form
\begin{eqnarray}
n(\bc)&=&
 \sum_{i<j<k<\ell}\Big\{c_{ij}c_{k\ell}(1-c_{i\ell})(1-c_{jk})+c_{i\ell}c_{jk}(1-c_{ij})(1-c_{k\ell})
\nonumber\\[-2mm]
&&
\hspace*{11mm}
+~c_{ij}c_{k\ell}(1-c_{ik})(1-c_{j\ell})+c_{ik}c_{j\ell}(1-c_{ij})(1-c_{k\ell})
\nonumber
\\
&&
\hspace*{11mm}
+~c_{ik}c_{j\ell}(1-c_{i\ell})(1-c_{jk})+c_{i\ell}c_{jk}(1-c_{ik})(1-c_{j\ell})
\Big\}
\nonumber
\\
&=&
\frac{1}{12}\sum_{ijk\ell}
\notdelta_{ik}\notdelta_{i\ell}\notdelta_{jk}\notdelta_{j\ell}
c_{ij}c_{k\ell}(1-c_{i\ell})(1-c_{jk})
+\frac{1}{12}\sum_{ijk\ell}
\notdelta_{ij}\notdelta_{ik}\notdelta_{j\ell}\notdelta_{k\ell}c_{i\ell}c_{jk}(1-c_{ij})(1-c_{k\ell})
\nonumber\\
&&
\hspace*{40mm}
+\frac{1}{12}\sum_{ijk\ell}\notdelta_{ij}\notdelta_{i\ell}\notdelta_{jk}\notdelta_{k\ell}c_{ik}c_{j\ell}(1-c_{i\ell})(1-c_{jk})
\nonumber
\end{eqnarray}
\clearpage
\begin{eqnarray}
&=&
\frac{1}{12}\sum_{ijk\ell}
(1-\delta_{ik}-\delta_{i\ell}-\delta_{jk}-\delta_{j\ell}+\delta_{ik}\delta_{j\ell}+\delta_{i\ell}\delta_{jk})
c_{ij}c_{k\ell}(1-c_{i\ell})(1-c_{jk})
\nonumber
\\
&&
+\frac{1}{12}\sum_{ijk\ell}
(1-\delta_{ij}-\delta_{ik}-\delta_{j\ell}-\delta_{k\ell}+\delta_{ij}\delta_{k\ell}+\delta_{ik}\delta_{j\ell})c_{i\ell}c_{jk}(1-c_{ij})(1-c_{k\ell})
\nonumber\\
\hspace*{-10mm}
&&
+\frac{1}{12}\sum_{ijk\ell}(1-\delta_{ij}-\delta_{i\ell}-\delta_{jk}-\delta_{k\ell}+\delta_{ij}\delta_{k\ell}+\delta_{i\ell}\delta_{jk})c_{ik}c_{j\ell}(1-c_{i\ell})(1-c_{jk})
\nonumber
\\
&=&
\frac{1}{4}\sum_{ijk\ell}c_{ij}c_{k\ell}(1-c_{i\ell})(1-c_{jk})+\frac{1}{4}\sum_{ij}c_{ij}
-\frac{1}{2}\sum_{ijk}
c_{ij}c_{ik}(1-c_{jk})
\nonumber
\\
&=&
\frac{1}{4}\Big(\sum_i k_i\Big)^2 +\frac{1}{4}\sum_{i}k_i
-\frac{1}{2}\sum_{i}k_i^2
-\frac{1}{2}\sum_{ij}k_{i}c_{ij}k_j
+\frac{1}{4}{\rm Tr}(\bc^4)+\frac{1}{2}{\rm Tr}(\bc^3)
\label{eq:nc}
\end{eqnarray}
where $k_i=\sum_j c_{ij}$ is the degree of node $i$.

Formula  (\ref{eq:nc}) plays a key role in the construction of our controlled Markov chain\footnote{The occurrence in (\ref{eq:nc}) of the terms ${\rm Tr}(\bc^3)$ and ${\rm Tr}(\bc^4)$ imply a strong connection between a graph's mobility $n(\bc)$ and its loop statistics, and it might be instructive to pursue this link somewhat further. If we call $L_n(\bc)$ the number of loops of length $n$ in $\bc$, one finds upon correcting for over-counting,
that $L_3(\bc)=\frac{1}{6}\sum_{ijk}c_{ij}c_{jk}c_{ki}=\frac{1}{6}{\rm Tr}(\bc^3)$ and that $L_4(\bc)=\frac{1}{8}\sum_{ijk\ell}(1-\delta_{j\ell})(1-\delta_{ik})c_{ij}c_{jk}c_{k\ell}c_{\ell i}=\frac{1}{8}{\rm Tr}(\bc^4)-\frac{1}{4}\sum_i k_i^2+\frac{1}{8}\sum_i k_i$. Hence we may write (\ref{eq:nc}) alternatively as
$n(\bc)=\frac{1}{4}(\sum_i k_i)^2-\frac{1}{2}\sum_{ij}k_i c_{ij}k_j +2 L_4(\bc)+3L_3(\bc)$.}, since we have seen earlier
that any dependence of $n(\bc)$ on the state $\bc$ that cannot be expressed in terms of the degree sequence only  will generate
entropic preferences of certain graphs over others. In appendix \ref{app:nc_bounds} we extract from (\ref{eq:nc})  the following rigorous bounds for $n(\bc)$, with $k_{\rm max}=\max_i k_i$ and $\bra k^m\ket=N^{-1}\sum_i k_i^m$:
\begin{eqnarray}
n(\bc)&\geq & \frac{1}{4}N^2\bra k\ket^2 +\frac{1}{4}N\bra k\ket
-\frac{1}{4}N\bra k^2\ket( 2k_{\rm max}+1)
\label{eq:lower}
\\
n(\bc)&\leq &
\frac{1}{4}N^2\bra k\ket^2 +\frac{1}{4}N\bra k\ket
-\frac{1}{4}N\bra k^2\ket
\label{eq:upper}
\end{eqnarray}
Let us test the full expression (\ref{eq:nc}) and the strength of the two bounds (\ref{eq:lower},\ref{eq:upper}) against three qualitatively different but explicitly verifiable cases:

\begin{itemize}
\item
Fully connected graphs
$c_{ij}=1-\delta_{ij}$:
\\[2mm]
Here one should find $n(\bc)=0$ (since no edge swaps are possible).
The degrees of $\bc$ are $k_i=N\!-\!1$ for all $i$, and its eigenvalues are $\lambda=N\!-\!1$ (with multiplicity 1) and $\lambda=\!-\!1$ (with multiplicity $N\!-\!1$). Hence ${\rm Tr}(\bc^4)=(N\!-\!1)[(N\!-\!1)^3+1]$ and ${\rm Tr}(\bc^3)=N(N\!-\!1)(N\!-\!2)$, and substitution of these properties  shows that (\ref{eq:nc})  indeed gives correctly $n(\bc)=0$. In this example the bounds (\ref{eq:lower},\ref{eq:upper}) are found to be weak, reducing to
$\frac{1}{4}N(N\!-\!1)[1\!-\!(N\!-\!1)^2]\leq n(\bc)\leq \frac{1}{4}N(N\!-\!1)[1\!+\!(N\!-\!1)^2]$.
\vsp

\item
Periodic chains $c_{ij}=\delta_{i,j-1}+\delta_{i,j+1}$ (mod $N$), $N\geq 4$:
\\[2mm]
 Direct inspection of the possible edge swaps reveals that one should find $n(\bc)=N(N\!-\!4)$.
In this ring-type graph $k_i=2$ for all $i$, $(\bc^3)_{ij}=\delta_{i,j+3}+3\delta_{i,j+1}+3\delta_{i,j-1}+\delta_{i,j-3}$ and
$(\bc^4)_{ij}=\delta_{i,j+4}+4\delta_{i,j+2}+6\delta_{ij}+4\delta_{i,j-2}+\delta_{i,j-4}$. It follows that
${\rm Tr}(\bc^4)=6N$ and ${\rm Tr}(\bc^3)=0$, and our formula (\ref{eq:nc}) is seen to reproduce
 $n(\bc)=N(N\!-\!4)$ correctly. The bounds (\ref{eq:lower},\ref{eq:upper}) are now rather close, giving
 $N(N\!-\!\frac{9}{2}) \leq n(\bc)\leq N(N\!-\!\frac{1}{2})$.
In fact one obtains the same leading two orders in $N$ of this result for $n(\bc)$ also in regular random graphs with $p(k)=\delta_{k,2}$, where the eigenvalue distribution of the adjacency matrix is \cite{Doro03}
\begin{equation}
\lim_{N\to\infty}\varrho(\lambda)=\frac{1}{\pi}\frac{\theta(2-|\lambda|)}{\sqrt{4-\lambda^2}}
\end{equation}
here one finds
\begin{eqnarray}
n(\bc)&=& N^2 -\frac{11}{2}N+\frac{1}{4}N\int\!\rmd \lambda~\varrho(\lambda)[\lambda^4+2\lambda^3]+{\sl o}(N)
\nonumber
\\
&=& N^2 -\frac{11}{2}N+8N\int_{0}^1\!\frac{\rmd x}{\pi}\frac{x^4}{\sqrt{1\!-\!x^2}}
+{\sl o}(N)
=
N(N-4)+{\sl o}(N)
\end{eqnarray}
 \vsp

\item
Two isolated links: $c_{12}=c_{21}=c_{34}=c_{43}=1$, with all other $c_{ij}=0$:
\\[2mm]
Here one should get $n(\bc)=2$. The degrees of $\bc$ are  $k_1=k_2=k_3=k_4=1$, with $k_i=0$ for $i>4$.
The eigenvalues are $\lambda=\pm 1$ (each with multiplicity 2) and $\lambda=0$ (with multiplicity $N\!-\!4$), so ${\rm Tr}(\bc^4)=4$ and ${\rm Tr}(\bc^3)=0$.  Again our formula (\ref{eq:nc}) is confirmed to be correct, reducing to $n(\bc)=2$.
The bounds (\ref{eq:lower},\ref{eq:upper}) now give $2\leq n(\bc)\leq 4$; here the lower bound is satisfied with equality.
\vsp
\end{itemize}

\noindent
It is not generally possible to simplify our formula (\ref{eq:nc}) for $n(\bc)$ further in
terms of the degree sequence only (had it been possible,  $n(\bc)$ would have dropped out of the transition probabilities (\ref{eq:FinalTransitionProbabilities})), as this would require that the third and fourth moments of the eigenvalue distribution $\varrho(\lambda)$ of any graph
can be written in terms of its degree sequence. Indeed, below we will discuss several examples of graphs with the same degree sequence that have different mobilities.

%%%%%%%%%%%%%%%%%%%%%%%

\subsection{Markov chain transition probabilities for edge swaps}

Our next task is to simplify the general expression (\ref{eq:FinalTransitionProbabilities}) for the transition probabilities of the appropriate Markov chain, for the case where the set of moves $\Phi^\prime$ is defined as above. We have
\begin{eqnarray}
W(\bc|\bc^\prime)&=& \sum_{i<j<k<\ell}~\sum_{\alpha\leq 3} \frac{I_{ijk\ell;\alpha}(\bc^\prime)}{n(\bc^\prime)}
\Big[
\frac{\delta_{\bc,F_{ijk\ell;\alpha}\bc^\prime} \rme^{-\frac{1}{2}[E(F_{ijk\ell;\alpha}\bc^\prime)-E(\bc^\prime)]}
+\delta_{\bc,\bc^\prime} \rme^{\frac{1}{2}[E(F_{ijk\ell;\alpha}\bc^\prime)-E(\bc^\prime)]}}{
\rme^{-\frac{1}{2}[E(F_{ijk\ell;\alpha}\bc^\prime)-E(\bc^\prime)]}+
\rme^{\frac{1}{2}[E(F_{ijk\ell;\alpha}\bc^\prime)-E(\bc^\prime)]}}
\Big]
\nonumber
\\[-2mm]
\hspace*{-21mm}
\end{eqnarray}
where $E(\bc)=H(\bc)+\log n(\bc)$. We can also write this as
\begin{eqnarray}
\bc\neq \bc^\prime:&~~~
W(\bc|\bc^\prime)&= \sum_{i<j<k<\ell}~\sum_{\alpha\leq 3} \frac{I_{ijk\ell;\alpha}(\bc^\prime)}{n(\bc^\prime)}
\frac{\delta_{\bc,F_{ijk\ell;\alpha}\bc^\prime}}{
1+\rme^{\Delta_{ijk\ell;\alpha}E(\bc^\prime)}}
\label{eq:edgeswapMC1}
\\
\bc=\bc^\prime:&~~~W(\bc^\prime|\bc^\prime)&=1-\sum_{\bc\neq \bc^\prime}W(\bc|\bc^\prime)
\label{eq:edgeswapMC2}
\end{eqnarray}
with
\begin{eqnarray}
\hspace*{-15mm}
\Delta_{ijk\ell;\alpha}E(\bc)&=&E(F_{ijk\ell;\alpha}\bc)-E(\bc)\nonumber
\\
\hspace*{-15mm}
&=& \Delta_{ijk\ell;\alpha}H(\bc) +\log\Big[1+ \Delta_{ijk\ell;\alpha}n(\bc)/n(\bc)\Big]
\label{eq:flipenergies}
\end{eqnarray}
Whether or not the entropic effects (the dependence of the mobility $n(\bc)$ on the graph state $\bc$) remain important in controlling the evolution of large graphs, i.e. for $N\to\infty$, will depend crucially on the degree distribution and the spectral properties of $\bc$. They can only be
neglected if the relative changes $\Delta_{ijk\ell;\alpha}n(\bc)/n(\bc)$ in the number of possible moves due to a single edge swap are always small.  It is therefore important to evaluate the change in the state mobility after an edge swap.

Only in the simplest case where all degrees are bounded, and the lowest moments of the degree distribution and of the eigenvalue spectrum $\varrho(\lambda)$ of
$\bc$ remain finite for $N\to\infty$, we would get
\begin{eqnarray}
\Delta_{ijk\ell;\alpha}E(\bc)&=&
\Delta_{ijk\ell;\alpha}H(\bc) +\frac{1}{N^2\bra k\ket^2}\Delta_{ijk\ell;\alpha}{\rm Tr}(\bc^4\!\!+\!2\bc^3)
 +\order(N^{-2})
 \nonumber
 \\
 &=&
 \Delta_{ijk\ell;\alpha}H(\bc) +~
 \order(N^{-1})
\end{eqnarray}
However, for graphs with e.g. $\Delta_{ijk\ell;\alpha}{\rm Tr}(\bc^4)=\order(N^2)$ it is less clear when these entropic effects can
be neglected. Imagine, for instance, a graph with a `dense core', such as $c_{ij}=1-\delta_{ij}$ for $i,j\in\{1,\ldots,K\}$ and $c_{ij}=0$ elsewhere.
Here one has $k_i=K-1$ for $i\in\{1,\ldots,K\}$ and $k_i=0$ for $i>K$ so $n(\bc)=0$, since this specific state cannot be changed by any edge swap. In addition, $\varrho(\lambda)=N^{-1}[\delta(\lambda\!-\!K\!+\!1)+(K\!-\!1)\delta(\lambda\!+\!1)+(N\!-\!K)\delta(\lambda)]$, which gives
\begin{eqnarray}
\hspace*{-10mm}
\bra k\ket^2+\frac{1}{N}\bra k\!-\!2k^2\ket=
-\frac{1}{N^{2}}\Big[
{\rm Tr}(\bc^4\!\!+\!2\bc^3)-2\sum_{rs}k_{r} c_{rs}k_s
\Big]=
\frac{K(K\!-\!1)(K^2\!-3K\!+\!3)}{N^2}
\end{eqnarray}
This shows that the $\order(N^2)$ terms in $n(\bc)$ cancel (in this example even the subsequent orders do), and as a result the numerator and denominator of $\Delta_{ijk\ell;\alpha}n(\bc)/n(\bc)$ are of the same order. Upon choosing $K=\order(\sqrt{N})$, for instance, one would have a
 finite nonzero average connectivity $\bra k\ket$ for $N\to\infty$,  but a diverging value of $\int\!\rmd\lambda~\lambda^4\varrho(\lambda)$ such that
 $N^{-2}{\rm Tr}(\bc^4)=\order(1)$.
   We must therefore allow for the possibility that
the entropic contribution in (\ref{eq:flipenergies}) will remain non-negligible
if one allows for small deviations from the above `dense core' graphs such as to render them mobile via edge swaps.

%%%%%%%%%%%%%%%%%%

\subsection{Ergodicity}

\noindent
The question of ergodicity for switching dynamics, i.e. whether any two graphs sharing a given degree sequence can be connected by a finite number of consecutive switchings, has been studied by Taylor \cite{Taylor}, where a formal proof by induction is given. Taylor's arguments focus on connected graphs\footnote{Actually, in \cite{Taylor} it is shown (Theorems 3.1 and 3.2) that ergodicity holds for generic connected pseudo-graphs, a class of structures that allows for multiple links between edges and for loops of length 1, which is much broader than just the simple graphs we consider here. The present case is covered by Theorem 3.3 in \cite{Taylor}.}, and this represents the harder case. Indeed, ergodicity for connected graphs would immediately transfer to generic graphs by a simple argument. Given a non-connected graph, add an auxiliary node connected to every other node. This extra node creates a new, connected graph. However the new links can never be switched with any of the pre-existing links. Hence they are not modified by the dynamics. Therefore, if ergodicity holds for the auxiliary connected graph, it is automatically valid also for the original non-connected graph. A more rigorous proof of the latter fact can be found in e.g. \cite{Egg}.

%%%%%%%%%%%%%%%%%%%%

\section{Invariant measure for edge swap dynamics with uniform acceptance probabilities}

Having established the precise connection between controlled stochastic
switching dynamics and the resulting equilibrium measures, we can use this in two ways. First, we can construct for any desired equilibrium measure a canonical edge-swap Monte-Carlo process that will evolve towards it. Alternatively, we can investigate edge-swap processes used by others,  and calculate the equilibrium states that would be generated. The simplest such processes are those where at each step a candidate edge swap is drawn randomly and uniformly and, when possible, executed. These are studied in this section.

\subsection{Relative errors in observables upon assuming incorrectly a uniform measure}

 An edge swap graph shuffling process in which all randomly drawn candidate moves are accepted and executed
 is mathematically equivalent (apart from an overall time re-scaling) to our general Monte-Carlo process
with transition probabilities (\ref{eq:edgeswapMC1},\ref{eq:edgeswapMC2},\ref{eq:flipenergies}), if in our formul\ae~ we make the choice $H(\bc)=-\log n(\bc)$. Since the general process (\ref{eq:edgeswapMC1},\ref{eq:edgeswapMC2},\ref{eq:flipenergies}) has been constructed such as to evolve to the equilibrium measure $p_\infty(\bc)=Z^{-1}\exp[-H(\bc)]$ we know that the one where all randomly drawn allowed edge swaps are accepted, must evolve to the stationary measure $p_\infty(\bc)=Z^{-1}\exp[\log n(\bc)]=Z^{-1}n(\bc)$, i.e. to
\begin{eqnarray}
p_\infty(\bc)&=& \frac{1}{Z}\Big\{
\bra  k\ket^2\! +\frac{1}{N}\Big(\bra k\ket
\!-\!2\bra k^2\ket
\!-\!\frac{2}{N}\!\sum_{ij}k_{i}c_{ij}k_j\Big)
+\frac{1}{N^{2}}{\rm Tr}(\bc^4+2\bc^3)\Big\}
\end{eqnarray}
where $Z$ is determined by the normalization condition $\sum_{\bc\in\Omega[\bk]}p_\infty(\bc)=1$.
We can now express the actual equilibrium expectation values $\bra G(\bc)\ket=\sum_{\bc\in\Omega[\bk]}p_\infty(\bc)G(\bc)$ of graph observables
in terms of what would have been found for a strictly flat measure, i.e. in terms of averages of the type $\bra G(\bc)\ket_0=|\Omega[\bk]|^{-1}\sum_{\bc\in\Omega[\bk]}G(\bc)$, where all graphs with the prescribed degree sequence $\bk$ having equal weight:
\begin{eqnarray}
\frac{\bra G(\bc)\ket-\bra G(\bc)\ket_0}{\bra G(\bc)\ket_0}&=& \frac{\sum_{\bc\in\Omega[\bk]}n(\bc)G(\bc)}{\sum_{\bc\in\Omega[\bk]}n(\bc)}
\frac{1}{\bra G(\bc)\ket_0}-1
\nonumber
\\
&=& \frac{\bra n(\bc)G(\bc)\ket_0-\bra n(\bc)\ket_0\bra G(\bc)\ket_0}{\bra n(\bc)\ket_0\bra G(\bc)\ket_0}
\nonumber
\\
&=&
 \frac{
\frac{\bra G(\bc)
{\rm Tr}(\bc^4+2\bc^3)\ket_0}{\bra G(\bc)\ket_0}
-\bra
{\rm Tr}(\bc^4\!+\!2\bc^3)\ket_0
-2\sum_{ij}k_ik_j\big[
\frac{\bra G(\bc)c_{ij}\ket_0}{\bra G(\bc)\ket_0}-\bra c_{ij}\ket_0\big]
}
{N^2
\bra  k\ket^2 +N[\bra k\ket
\!-\!2\bra k^2\ket]\big]
+\bra
{\rm Tr}(\bc^4\!+\!2\bc^3)\ket_0 -2\sum_{ij}k_{i}k_j\bra c_{ij}\ket_0}
\label{eq:deviations_from_uniform}
\end{eqnarray}
According to (\ref{eq:FinalAcceptanceRates}), the correct dynamics leading to the uniform measure $p(\bc)=|\Omega[\bk]|^{-1}$ would have corresponded
to the following acceptance probabilities for proposed edge swaps $\bc^\prime \to\bc$:
\begin{eqnarray}
A(\bc|\bc^\prime)&=& n(\bc^\prime)/[n(\bc^\prime)+n(\bc)]
\label{eq:acceptance_flat}
\end{eqnarray}
 A sensitive marker of deviations from uniform sampling of graphs by $p_\infty(\bc)$ should be the observable $n(\bc)$ itself (\ref{eq:nc}), for which one finds
the relative error
\begin{eqnarray}
\frac{\bra n(\bc)\ket-\bra n(\bc)\ket_0}{\bra n(\bc)\ket_0}
&=& \frac{\bra n^2(\bc)\ket_0-\bra n(\bc)\ket_0^2}{\bra n(\bc)\ket_0^2}\geq 0
\end{eqnarray}
with equality if and only if $n(\bc)=n(\bc^\prime)$ for all $\bc,\bc^\prime\in\Omega[\bk]$, i.e. if the measure $p_\infty(\bc)$ is flat.
Let us next acquire some intuition for the degree sequences $\bk=(k_1,\ldots,k_N)$ where the incorrect assumption of a uniform equilibrium
measure over $\Omega[\bk]$ would lead to significant errors in expectation values. These situations occur when the mobilities $n(\bc)$ vary significantly from one graph to another.

%%%%%%%%%%%%%%%%%

\subsection{`Nearly hardcore' graphs}

\begin{figure}[t]
\vspace*{0mm} \hspace*{-10mm} \setlength{\unitlength}{0.7mm}
\begin{picture}(200,100)

  \put(0,0){\includegraphics[height=110\unitlength]{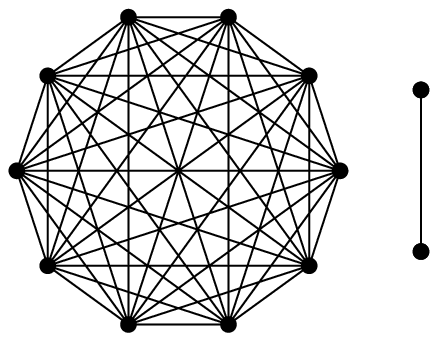}}
 \put(97,65){$K\!+\!1$} \put(97,40){$K\!+\!2$}
 \put(50,95){type A}

  \put(120,0){\includegraphics[height=110\unitlength]{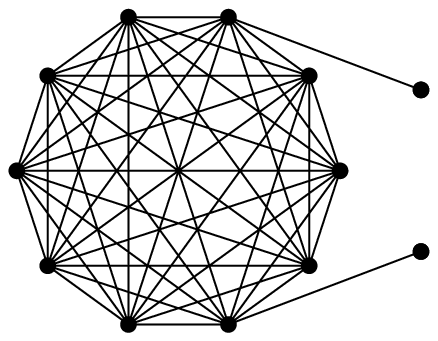}}
 \put(217,65){$K\!+\!1$} \put(217,40){$K\!+\!2$}
 \put(185,82){\large$k$} \put(185,23){\large$\ell$}
 \put(170,95){type B}

\end{picture}
 \vspace*{-15mm}
\caption{Connected parts of `nearly hard-core' graphs, as defined by a degree sequence (modulo node permutations)
of the form: $k_i=K-1$ for $i\leq K$, $k_i=1$ for $i\in\{K+1,K+2\}$, and $k_i=0$ for $i>K+2$.
In the present figure  $K=10$, and the
nodes $i>K+2$ with zero degree are not shown.
Left: the graph of type A, with $c_{K+1,K+2}=1$, of which there is only one; here $n(\bc)=K(K-1)$.
Right: the graphs of type B, with $c_{K+1,K+2}=0$, of which there are $K(K-1)$ (one for each choice of $k$ and $\ell$); here $n(\bc)=2(K-1)$. An equilibrated stochastic edge swap dynamics with randomly drawn candidate edge swaps and state-independent acceptance rates would
visit the graph A with probability $p(\bc)=1/[1+2(K-1)]$, and each of the $K(K-1)$
type B graphs  with probability $p(\bc)=2/[K(1+2(K-1))]$.
\label{nearly}}
\label{fig:softcore}
\end{figure}

As an example let us consider a choice for the degrees $\bk=(k_1,\ldots,k_N)$ that corresponds to a `nearly hardcore' graph (see Fig. \ref{nearly}), involving a fully connected core of size $K$ and two degree-1 nodes (note that for a fully `hardcore' graph with $K=N$ there would not be any allowed edge swap):
\begin{eqnarray}
&&i\leq K:~~ k_i=K\!-\!1,~~~~~i=K\!+\!1,K\!+\!2:~~k_i=1,~~~~~i>K\!+\!2:~~k_i=0
\end{eqnarray}
Here we have $p(k)=\frac{K}{N}\delta_{k,K-1}+\frac{2}{N}\delta_{k,1}+\frac{N-K-2}{N}\delta_{k,0}$,
and hence formula (\ref{eq:nc}) gives
\begin{eqnarray}
4n(\bc)&=&
K^2(K\!-\!1)^2+5K(K\!-\!1)-2K(K\!-\!1)^2+2+{\rm Tr}(\bc^4+2\bc^3)
\nonumber
\\
&& - 2(K\!-\!1)^2\sum_{ij=1}^{K}c_{ij}-4(K\!-\!1)\sum_{i=1}^K [c_{i,K+1}+c_{i,K+2}]-4c_{K+1,K+2}
\label{eq:nc_example}
\end{eqnarray}
There exist $K(K\!-\!1)+1$ such graphs. Close inspection of their possible realizations reveals only two types, A and B, which are characterized by whether or not the two degree-1 nodes are connected to each other, see Figure \ref{fig:softcore}:
\vsp

\begin{itemize}
\item Type A:
\begin{eqnarray}
c_{K+1,K+2}=1,~~~~~~c_{ij}=1~~\forall (i,j),~i<j\leq K,~~~~~~
c_{ij}=0~~{\rm elsewhere}
\end{eqnarray}
Here we have the two degree-1 nodes are connected to each other, plus one fully connected core of $K$ nodes.
There is just one such type-A graph $\bc$, which we will call $\bc_{\rm A}$.
 It allows only for edge swaps involving the two degree-1 nodes and any two nodes taken from the core, so $n(\bc)=K(K-1)$. The eigenvalue distribution of $\bc$ is
\begin{eqnarray}
\varrho(\lambda)&=& \frac{1}{N}\delta(\lambda\!-\!1)+\frac{K}{N}\delta(\lambda\!+\!1)+\frac{1}{N}\delta(\lambda\!-\!K\!+\!1)+\frac{N\!-\!K\!-\!2}{N}\delta(\lambda)
\end{eqnarray}
and hence formula (\ref{eq:nc_example}) indeed reproduces the correct mobility $n(\bc)=K(K-1)$.
\vsp

\item Type B:
\begin{eqnarray}
 c_{K+1,K+2}=0,~~~~~~\exists(k,\ell \leq K)~{\rm such~that}&~& c_{k,K+1}=c_{\ell,K+2}=1,
\nonumber \\
                  &&c_{ij}=1~~\forall (i,j),~ i\!<\!j\!\leq\! K,~i,j\notin \{k,\ell\}  \\
                  &&c_{ij}=0~~{\rm elsewhere}
                  \nonumber
\end{eqnarray}
 Here we have de facto carried out an edge swap relative to $\bc_{\rm A}$, and replaced both the link between the two degree-1 nodes and the link between two nodes $(k,\ell)$ of the core by two new links: one from node $k$ to degree-1 node $K+1$, and one from node $\ell$ to the other degree-1 node $K+2$. We will call the resulting graph $\bc_{{\rm B};k,\ell}$.
There are $K(K-1)$ such type-B graphs, one for each choice of $(k,\ell)$, but each of these allows only for $2(K-1)$ edge swaps. In fact, the two outlying edges can be switched among each other in two ways, the one that brings us back to $\bc_{\rm A}$ and the swap that is equivalent to replacing the links $c_{k,K+1}$ and $c_{\ell,K+2}$ by the links $c_{k,K+2}$ and $c_{\ell,K+1}$. In addition, each of the outlying edges can be swapped separately in one way with every edge that has a node in common with the other outlying edge (this is easily checked directly). There are $K-2$ such edges. Hence $n(\bc)=2+2(K-2)=2(K-1)$ for each state of type B. The spectrum of each type-B state is
\begin{eqnarray}
\varrho(\lambda)&=& \frac{1}{N}\delta(\lambda\!-\!1)+\frac{1}{N}\delta(\lambda\!+\!1)+\frac{K}{N}\tilde{\varrho}(\lambda)+\frac{N\!-\!K\!-\!2}{N}\delta(\lambda)
\end{eqnarray}
in which the normalized distribution $\tilde{\varrho}(\lambda)$ represents an eigenspace of dimension $K$ with three further eigenvalues, to be solved from
$(\lambda+1)(\lambda^2-1)=(K-2)(\lambda^2+2\lambda-1)$.
\end{itemize}
\vsp

\noindent
We see explicitly that here the edge swap dynamics is indeed ergodic on the space of allowed graphs.
It follows from the possible graphs $\bc$ and from their associated mobility numbers $n(\bc)$ as identified  above, that for the present choice of degrees one will find under edge swap dynamics with state-independent acceptance probabilities the following equilibrium measure:
\begin{eqnarray}
p_\infty(\bc)=\frac{1}{1+2(K-1)}\delta_{\bc,\bc_{\rm A}}+\frac{2}{K[1+2(K-1)]}\sum_{k,\ell=1}^K\notdelta_{k\ell}\delta_{\bc,\bc_{{\rm B};k,\ell}}
\label{eq:measure_softcore}
\end{eqnarray}
This measure is far from uniform. It will now depend crucially on which quantity one measures during the graph evolution process to which extent this probability inhomogeneity will manifest itself. If we were to measure, for instance, the expectation value of the link $c_{K+1,K+2}$ (which is only present in the state $\bc_{\rm A}$) we would find from the true equilibrium measure (\ref{eq:measure_softcore}):
\begin{eqnarray}
\bra c_{K+1,K+2}\ket &=& [1+2(K-1)]^{-1}
\label{eq:false}
\end{eqnarray}
whereas for the uniform measure $p(\bc)=|\Omega[\bk]|^{-1}=[K(K\!-\!1)\!+\!1]^{-1}$ one would have found:
\begin{eqnarray}
\bra c_{K+1,K+2}\ket_0= [1\!+\!K(K\!-\!1)]^{-1}
\label{eq:true}
\end{eqnarray}
Expressions (\ref{eq:false}) and (\ref{eq:true}) are identical only for $K=2$ (where the `dense core' is indeed no longer present, and the graph  reduces to two disconnected links), and differ significantly for increasing $K$:
\begin{eqnarray}
\frac{\bra c_{K+1,K+2}\ket-\bra c_{K+1,K+2}\ket_0}{\bra c_{K+1,K+2}\ket_0}&=&
\frac{(K-2)(K-1)}{1+2(K-1)}
\end{eqnarray}
Assuming (incorrectly) a uniform equilibrium measure on $\Omega[\bk]$
would grossly underestimate the true likelihood to observe the bond $c_{K+1,K+2}$.
Similarly, if we calculate for both (\ref{eq:measure_softcore}) (obtained for `accept all' implementation of randomly proposed edge swaps) and for the uniform measure on $\Omega[\bk]$ (obtained for the correct acceptance rates (\ref{eq:acceptance_flat})) the average mobilities $\bra n(\bc)\ket$ and $\bra n(\bc)\ket_0$, respectively, we find
\begin{eqnarray}
&&
\bra n(\bc)\ket= \frac{(K-1)(5K-4)}{2K-1},~~~~~~\bra n(\bc)\ket_0=\frac{K(K-1)(2K-1)}{K(K-1)+1}
\label{eq:compare_nhc}
\end{eqnarray}
Again these are identical only when $K=2$, whereas $\bra n(\bc)\ket>\bra n(\bc)\ket_0$ in all other cases.

Degree statistics such as in the present example do not strike us as far-fetched. If we were to add to
the present $K\!+\!2$ nodes another $N\!-\!K\!-\!2$ `dummy' nodes with degree zero, then our enlarged graph would have
$\bra k\ket=[K(K\!-\!1)\!+\!2]/N$ and $\bra k^2\ket=[K(K\!-\!1)^2\!+\!2]/N$. Choosing e.g. $K=\sqrt{\phi N}$ would then give a finite average connectivity
$\bra k\ket=\phi+\order(N^{-1/2})$
and a diverging width, $\bra k^2\ket=\order(N^{1/2})$, identical to what one would have found for scale-free graphs with $p(k)\sim k^{-5/2}$.
Nothing in the first two moments of the degree distribution could therefore be regarded as severely pathological.

%%%%%%%%%%%%%%%%

\subsection{Conditions under which the invariant measure will become uniform for $N\to\infty$}

It is clear from the above example that any results obtained from measuring graph observables numerically during edge swap dynamics with state-independent acceptance probabilities should be accompanied by an explicit proof that for the degree sequence under consideration and for the relevant graph size the equilibrium measure on the space of graphs with this given degree sequence can be taken as uniform.
According to (\ref{eq:edgeswapMC1},\ref{eq:edgeswapMC2},\ref{eq:flipenergies}), our stochastic
process will evolve towards a state with uniform graph probabilities on $\Omega[\bk]$ (the set of graphs characterized by an imposed degree sequence) if it corresponds to $H(\bc)={\rm constant}$ for all $\bc\in\Omega[\bk]$.
From this it follows that for $N\to\infty$ the canonical edge swap acceptance probabilities lose their dependence on $\bc$
as soon as
\begin{equation}\label{conda}
\lim_{N\to\infty}\Delta_{ijk\ell;\alpha}\log n(\bc)\equiv
\lim_{N\to\infty}\log\Big[1+\frac{\Delta_{ijk\ell;\alpha} n(\bc)}{ n(\bc)}\Big]=0
\end{equation}
In appendix \ref{app:nc_bounds} we prove bounds for $n(\bc)$ and $\Delta_{ijk\ell;\alpha} n(\bc)$. In particular, we may use (\ref{eq:lower_bound_nc}) and (\ref{eq:swap_bound_nc}), viz.
\begin{eqnarray}
&&
n(\bc)~\geq~ \frac{1}{4}N^2\bra k\ket^2 +\frac{1}{4}N\bra k\ket
-\frac{1}{2}N\bra k^2\ket( k_{\rm max}\!+\!\frac{1}{2})
\\
&&
|\Delta_{ijk\ell;\alpha}n(\bc)|~\leq~
\frac{1}{2}N\bra k^2\ket k_{\rm max}
\end{eqnarray}
to establish
\begin{eqnarray}
\frac{|\Delta_{ijk\ell;\alpha} n(\bc)|}{ n(\bc)}&\leq &
\frac{2\Lambda_N}{1-2\Lambda_N},~~~~~~~~\Lambda_N=
\frac{\bra k^2\ket (k_{\rm max}\!+\!\frac{1}{2})}{N\bra k\ket^2 }
\end{eqnarray}
This latter formula thus allows us to identify conditions under which the simple process with state-independent acceptance probabilities of randomly drawn candidate edge swaps will lead to an effectively\footnote{By `effectively uniform' we mean that the canonical edge swap acceptance
probabilities $1/[1+\rme^{\Delta_{ijk\ell;\alpha}E(\bc^\prime)}]$ in (\ref{eq:edgeswapMC1}) that would give us a rigorously uniform invariant measure
become for $N\to\infty$ indistinguishable from $1/2$ (i.e. from state-independent acceptance probabilities). This still does not rule out the possibility that certain pathological observables could be defined for which the asymptotically vanishing deviations from $1/2$ could add up to a non-vanishing effect.} uniform measure. Condition (\ref{conda}) will be satisfied if $\Lambda_N\ll 1$, i.e. as soon as
\begin{eqnarray}
\bra k^2\ket k_{\rm max}/\bra k\ket^2 \ll N
\label{eq:conditions_flat_measure}
\end{eqnarray}
Clearly, degree sequences of regular random graphs, viz. $p(k)=\delta_{k,k^\star}$ for some finite connectivity $k^\star$, meet all the conditions
for finding asymptotically a flat invariant measure. However, one has to be careful with scale-free degree sequences, where both $\bra k^2\ket$ and  $k_{\rm max}$ diverge as $N\to\infty$.

It is instructive to inspect condition (\ref{eq:conditions_flat_measure}) for the `nearly hardcore' graphs in the previous subsection, where
$p(k)=\frac{K}{N}\delta_{k,K-1}+\frac{2}{N}\delta_{k,1}+\frac{N-K-2}{N}\delta_{k,0}$, and where for state-independent acceptance rates one
does not have a flat invariant measure for any $K>2$. Here $\bra k\ket=K(K\!-\!1)/N+2/N$, $\bra k^2\ket=K(K\!-\!1)^2/N+2/N$ and $k_{\rm max}=K\!-\!1$, so
\begin{eqnarray}
\frac{\bra k^2\ket k_{\rm max}}{N\bra k\ket^2}&=& \frac{[K(K\!-\!1)^2+2](K\!-\!1)}{[K(K\!-\!1)+2]^2}
\label{eq:hardcore_ratio}
\end{eqnarray}
For all $K>2$ the ratio (\ref{eq:hardcore_ratio}) stays finite, so (\ref{eq:conditions_flat_measure}) is indeed violated.

Note that in the previous example also $K=2$ would violate (\ref{eq:conditions_flat_measure}), yet in this case one would have found a flat measure even for state independent acceptance rates.
It should therefore be emphasized that condition (\ref{eq:conditions_flat_measure}) has been derived as {\em sufficient} for having
effectively a uniform invariant measure. It is, however, not a {\em necessary} condition; there exist indeed degree sequences that violate
 (\ref{eq:conditions_flat_measure}) but still give a flat invariant measure. This is easily seen upon inspecting simple examples such as the `star-like' graphs. For instance, the graph characterized by the degrees $k_i=1$ for $i<N\!-\!4$, $k_{N-4}=k_{N-3}=k_{N-2}=k_{N-1}=2$, and $k_N=N-1$, corresponds to a central node $N$ connected to $N\!-\!5$ degree-1 nodes and connected to two further loops of length three.
This degree sequence has $\lim_{N\to\infty}\bra k\ket=2$, but $\bra k^2\ket =\order(N)$ and $k_{\rm max}=\order(N)$, and hence violates (\ref{eq:conditions_flat_measure}). Yet there are only three allowed graphs which each have $n(\bc)=2$, so the invariant measure is flat.
For some graphs it is in fact possible to sharpen the condition (\ref{eq:conditions_flat_measure}) further, by using alternative bounds for the various terms in $\Delta_{ijk\ell;\alpha}n(\bc)$; examples of these are derived at the end of appendix \ref{app:traces}.

%%%%%%%%%%%%%%%%%%%%

\section{Generating random graphs with prescribed degree correlations via edge swaps}

Our second application of the general formalism is the construction of Monte-Carlo processes that evolve towards the equilibrium state corresponding to
random graphs with prescribed degree sequences and controlled non-uniform measures. For instance, one could impose non-uniform measures in order to impose specific
degree correlations, which is achieved by the ensemble studied in \cite{PerezCoolen},
\begin{eqnarray}
p_\infty(\bc)&=& \frac{1}{Z}\prod_{i<j}\left[\frac{\bra k\ket
}{N}Q(k_i,k_j)\delta_{c_{ij},1}+\Big(1\!-\!\frac{\bra k\ket
}{N}Q(k_i,k_j)\Big)\delta_{c_{ij},0}\right]\prod_i\delta_{k_i,k_i(\bc)}
\label{eq:newer_connectivity}
 \end{eqnarray}
 with $Q(k,k^\prime)\geq 0$ for all $(k,k^\prime)$ and $\sum_{kk^\prime}p(k)p(k^\prime)Q(k,k^\prime)=1$.
This ensemble gives the maximum entropy within the subspace of graphs with prescribed degrees and upon imposing as a constraint the average values $\Pi(k,k^\prime)=\bra \Pi(k,k^\prime|\bc)\ket$ of the relative   degree correlations, where $\Pi(k,k^\prime|\bc)$ is defined as follows
\begin{eqnarray}
k,k^\prime >0:~~~
\Pi(k,k^\prime|\bc)&=& \frac{\sum_{i\neq j} c_{ij}~\delta_{k,k_i(\bc)}\delta_{k^\prime,k_j(\bc)}}
{ \sum_{i\neq j}\delta_{k,k_i(\bc)}\delta_{k^\prime,k_j(\bc)}}
\frac{\bra k\ket (N\!-\!1)}{kk^\prime}
\label{eq:Pi}
\end{eqnarray}
$\Pi(k,k^\prime|\bc)$ gives the probability that two randomly drawn nodes of $\bc$ with degrees $(k,k^\prime)$ are found to be connected, divided by the probability that this would be true in random graphs drawn from $p(\bc)=Z^{-1}\prod_i \delta_{k_i,k_i(\bc)}$. For the ensemble (\ref{eq:newer_connectivity})
one finds for $N\to\infty$ that $\Pi(k,k^\prime)=Q(k,k^\prime)/F(k)F(k^\prime)$, where $F(k)$ is to be solved from
\begin{eqnarray}
\forall k\geq 0:&~~~& \sum_{k^\prime}Q(k,k^\prime)p(k^\prime)k^\prime /F(k^\prime)=\bra k\ket F(k)
\label{eq:solvingF}
\end{eqnarray}
Conversely, for each desired function $\Pi(k,k^\prime)$ one may always choose $Q(k,k^\prime)=\Pi(k,k^\prime)kk^\prime/\bra k\ket^2$ in
(\ref{eq:newer_connectivity}) and find this ensemble subsequently generating graphs with the required degree correlations.
See \cite{Annibale_etal} for proofs of these statements, and for further mathematical properties of $\Pi(k,k^\prime)$ and the ensemble (\ref{eq:newer_connectivity}).

The controlled non-uniform measure (\ref{eq:newer_connectivity}) can be generated via stochastic processes as studied in the present paper.
 In the language of our processes (\ref{eq:edgeswapMC1},\ref{eq:edgeswapMC2},\ref{eq:flipenergies}),
it simply corresponds to the choice
 \begin{eqnarray}
H(\bc)&=&-\sum_{i<j}\log \Big[\frac{\bra k\ket
}{N}Q(k_i,k_j)\delta_{c_{ij},1}+\Big(1\!-\!\frac{\bra k\ket
}{N}Q(k_i,k_j)\Big)\delta_{c_{ij},0}\Big]
\label{eq:H_forPC}
\end{eqnarray}
If we now work out the implications for (\ref{eq:edgeswapMC1},\ref{eq:edgeswapMC2},\ref{eq:flipenergies}) of choosing (\ref{eq:H_forPC}), we find this process describing the random drawing of candidate edge swaps $F_{ijk\ell;\alpha}$, upon which the proposed transition $\bc\to \bc_{\rm new}=F_{ijk\ell;\alpha}\bc$ is then accepted (and executed) with the acceptance probability
\begin{eqnarray}
A(\bc_{\rm new}|\bc)&=& \left[1+\frac{n(\bc_{\rm new})}{n(\bc)}~e^{H(\bc_{\rm new})-H(\bc)}\right]^{-1}
\label{eq:PC_A}
\end{eqnarray}
With the sets $S_{ijk\ell;\alpha}$ introduced in (\ref{eq:S1},\ref{eq:S2},\ref{eq:S3}) that specify which index pairs are affected by the proposed edge swap,
we find for the function (\ref{eq:H_forPC}) that
\begin{eqnarray}
e^{H(\bc_{\rm new})-H(\bc)}&=& \prod_{(a,b)\in S_{ijk\ell;\alpha}}
\left[\frac{\frac{\bra k\ket
}{N}Q(k_a,k_b)c_{ab}+\Big(1\!-\!\frac{\bra k\ket}{N}Q(k_a,k_b)\Big)(1-c_{ab})}
{\frac{\bra k\ket
}{N}Q(k_a,k_b)c^{\rm new}_{ab}+\Big(1\!-\!\frac{\bra k\ket}{N}Q(k_a,k_b)\Big)(1-c^{\rm new}_{ab})}
\right]
\nonumber
\\
&=& \prod_{(a,b)\in S_{ijk\ell;\alpha}}
\Big[L_{ab}\delta_{c_{ab}^{\rm new},1}+L^{-1}_{ab}\delta_{c_{ab}^{\rm new},0}\Big]
\label{eq:PC_deltaH}
\end{eqnarray}
(where we used the property $c_{ab}=1-c_{ab}^{\rm new}$ for all $(a,b)\in S_{ijk\ell;\alpha}$), with
\begin{eqnarray}
L_{ab}&=& N/[\bra k\ket Q(k_a,k_b)]-1
\label{eq:PC_L}
\end{eqnarray}
If we start from a physically realizable function $\Pi(k,k^\prime)$ (see \cite{Annibale_etal} for the precisely mathematical conditions for realizability) and if we use the canonical kernel $Q(k,k^\prime)=\Pi(k,k^\prime)kk^\prime/\bra k\ket^2$
in our ensemble (such that for $N\to\infty$ it will generate graphs with relative degree correlations $\Pi(k,k^\prime)$), the latter parameters
become $L_{ab}=N\bra k\ket/[\Pi(k,k^\prime)kk^\prime] -1$.

\section{Numerical tests}

We have conducted extensive numerical experiments on a variety of graphs to confirm the validity of formula (\ref{eq:nc}) for $n(\bc)$ and found an impressive agreement. The calculation of the terms ${\rm Tr}(\bc^3)$ and ${\rm Tr}(\bc^4)$ in $n(\bc)$ at each time step, required in the calculation of the acceptance probabilities of the canonical Markov chain, is cpu-intensive. However, for finitely connected graphs working out these traces can generally still be done in $\order(N)$ steps by efficient use of arrays with indices of the neighbours of each node, as opposed to  brutal matrix multiplication. Alternatively one could calculate the traces only once at the start of the simulation, and update their values on-line by using formulas  (\ref{eq:Tr3term},\ref{eq:Tr4term}) in appendix \ref{app:traces}.

\subsection{Accept-all edge swap dynamics versus edge-swap dynamics with correct acceptance rates}

\begin{figure}[t]
\vspace*{0mm} \hspace*{30mm} \setlength{\unitlength}{0.7mm}
\begin{picture}(200,90)

  \put(0,0){\includegraphics[height=90\unitlength]{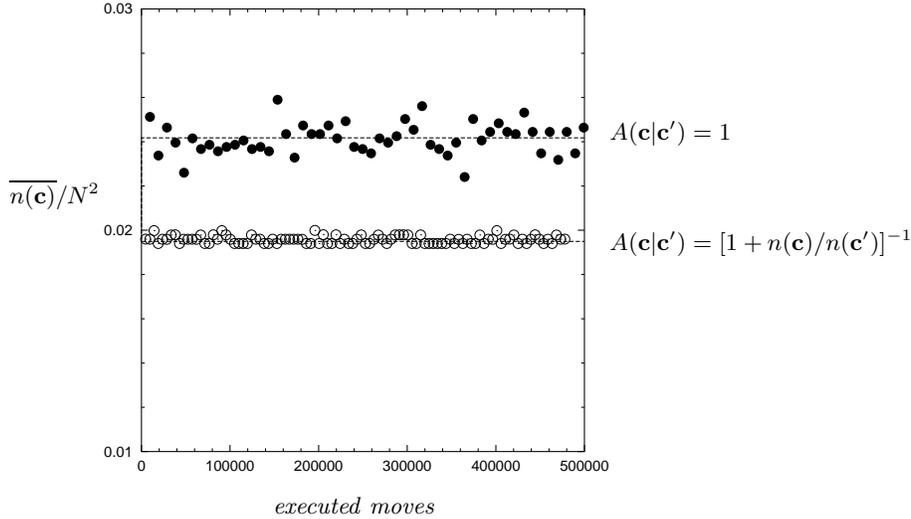}}
 \put(99,62){$A(\bc|\bc^\prime)=1$} \put(99,41){$A(\bc|\bc^\prime)=[1+n(\bc)/n(\bc^\prime)]^{-1}$}
 \put(-15,50){$\overline{n(\bc)}/N^2$}
 \put(35,-9){\em executed moves}

\end{picture}
 \vspace*{10mm}
\caption{Re-scaled running averages $\overline{n(\bc)}/N^2$ of the graph mobility (\ref{eq:nc}), measured during numerical simulations of Markov chains of the type (\ref{eq:edgeswapMC1},\ref{eq:edgeswapMC2}). The time unit is the number of executed edge swaps, and running averages are
measured over time windows of $10,\!000$ successive states $\bc$. All data refer to the nearly hardcore graphs shown in figure \ref{fig:softcore}, with $K=98$ and $N=100$. Full circles: observed graph mobility if {\em all} randomly proposed edge swaps that are possible are executed, i.e. when
$A(\bc|\bc^\prime)=1$. Open circles: observed graph mobility if
randomly proposed edge swaps that are possible are executed with the canonical acceptance rates (\ref{eq:acceptance_flat}), i.e. when $A(\bc|\bc^\prime)=[1+n(\bc)/n(\bc^\prime)]^{-1}$.
Dashed horizontal lines: the corresponding theoretical predictions (\ref{eq:compare_nhc}) for the equilibrium mobilities, which for $K=98$ give $\bra n(\bc)\ket/N^2\!\approx 0.0242$ (upper) and $\bra n(\bc)\ket_0/N^2\!\approx 0.0195$.
}
\label{fig:softcore_simulation}
\end{figure}

Our first simulations were carried out for the `nearly hardcore' graphs of figure \ref{fig:softcore}. We ran two different Monte-Carlo processes as described by (\ref{eq:edgeswapMC1},\ref{eq:edgeswapMC2}). In the first we accepted all randomly generated possible edge swaps, i.e. $A(\bc|\bc^\prime)=1$, whereas in the second process  we used the canonical acceptance rates (\ref{eq:acceptance_flat}), i.e.  $A(\bc|\bc^\prime)=[1+n(\bc)/n(\bc^\prime)]^{-1}$. This allowed us to verify the theoretical predictions that in the former process the system will evolve towards the non-uniform measure $p_\infty(\bc)=n(\bc)/\sum_{\bc^\prime\in\Omega[\bk]}n(\bc^\prime)$ whereas in the latter `the system evolves towards the flat measure $p_\infty(\bc)=|\Omega[\bk]|^{-1}$. This verification is easiest upon measuring time averages of the mobility $n(\bc)$ itself, for which we
have derived the (exact) expressions (\ref{eq:compare_nhc}). In figure \ref{fig:softcore_simulation} we show for both processes the observed mobility time averages $\overline{n(\bc)}$ as measured over successive time windows of $10^4$ accepted moves (which gives us also information on when the system can be regarded as in equilibrium), together with the predicted equilibrium values (\ref{eq:compare_nhc}) (as dashed horizontal lines).
We conclude that in these graphs there is perfect agreement between theory and the simulations, and that indeed one cannot generally assume `accept all' edge swap randomization to lead to an unbiased sampling of $\Omega[\bk]$.

\begin{figure}[t]
\vsp

\begin{center}
\includegraphics[height=6cm]{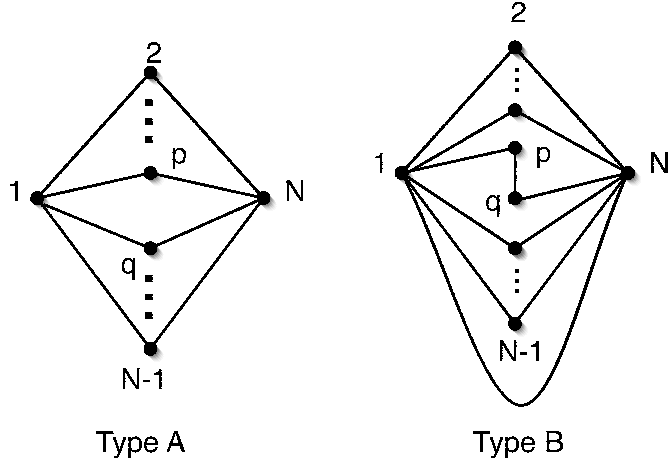}
\vspace*{5mm}
\caption{\label{milo2}Graphs used in our second numerical experiment (the undirected versions of the graphs introduced in \cite{milo2}). Left panel:  the graph chosen as the initial state of the Monte-Carlo process (\ref{eq:edgeswapMC1},\ref{eq:edgeswapMC2}) (to be called type A), with $K=N-2$ `central' nodes of degree 2 and two nodes of degree $K$. This configuration has $n(\bc)=K(K-1)$. A generic switching of a type A graph, e.g. one involving the four nodes $\{1,p,q,N\}$, leads to one of the $K(K-1)$ type B graphs, of which an example is shown in the right panel. Each of the type B graphs has $n(\bc)=2(K-1)$.}
\end{center}
\end{figure}

\begin{figure}[t]
\begin{center}
\includegraphics[height=6cm]{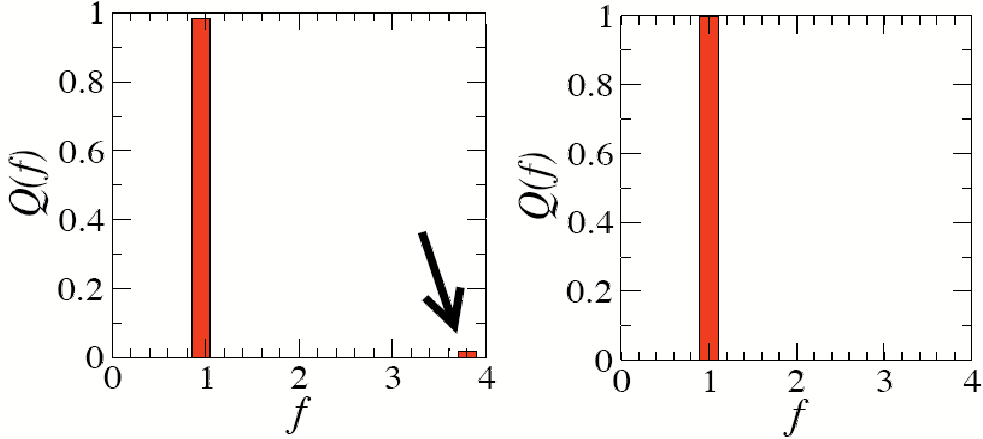}
\vspace*{5mm}
\caption{\label{milo3}
Distribution $Q(f)$ re-scaled frequencies at which the accessible graphs are visited during switching dynamics of the type (\ref{eq:edgeswapMC1},\ref{eq:edgeswapMC2}) over $2\cdot 10^6$ time steps, starting from a type A graph of figure \ref{milo2} with $N=10$. Here $|\Omega[\bk]|=57$. A uniform measure $p(\bc)=|\Omega[\bk]|^{-1}$  would give $Q(f)=\delta(f-1)$. In the left panel, we see the results for an `accept all' switching dynamics, viz. $A(\bc|\bc^\prime)=1$, with the arrow pointing to the peak corresponding to the time spent in the type A graph. In the right panel, the same quantity is reported for a switching dynamics with the canonical acceptance probabilities $A(\bc|\bc^\prime)=[1+n(\bc)/n(\bc^\prime)]^{-1}$.}
\end{center}
\end{figure}

As our second synthetic example system, we choose as our starting
point the undirected version of the graph studied in \cite{milo2}, shown in figure \ref{milo2}, to be called graph A (or $\bc_A$).
It consists of $N$ nodes, two of which (labeled $1$ and $N$) have degree $N-2$, whereas the remaining $N-2$ nodes have two connections each. We note that $n(\bc_A)=(N-2)(N-3)$, since it is possible to switch in one way every two links joining opposite sides of the central `wall' provided they don't have a node in common (this mobility value can of course also be calculated from (\ref{eq:nc})). Any possible edge swap executed on $\bc_A$ will bring us to a graph of type B, see right diagram in figure \ref{milo2}, of which there are $(N-2)(N-3)$. It follows that $|\Omega[\bk]|=1+(N-2)(N-3)$.
Looking at Fig. \ref{milo2}, we see that to compute the mobility of type B graphs one has to take into account the facts that: (i) we can switch in one way the links such as $(1,p)$ and $(q,N)$, and (ii) the link $(q,p)$ can be switched in one way with each of the links $(1,i)$ and $(j,N)$ such that $i\notin\{p,N\}$ and $j\notin\{1,q\}$. There are $N-4$ such links in each group. Hence including the switch that brings us back to type A we have $n(\bc_B)=1+2(N-4)+1=2(N-3)$ (again, one could also have used formula (\ref{eq:nc}) to find this result).
If one carries out an `accept all' edge swap process starting from $\bc_A$, our theory predicts that upon equilibration this would give
the following measure on the space $\Omega[\bk]$ of graphs with degrees sequences identical to that of $\bc_A$:
\begin{eqnarray}
p_\infty(\bc_A)&=&\frac{n(\bc_A)}{n(\bc_A)+(N\!-\!2)(N\!-\!3)n(\bc_B)}=\frac{1}{2N-5}
\end{eqnarray}
and for each type B graph
\begin{eqnarray}
p_\infty(\bc_B)&=&\frac{n(\bc_B)}{n(\bc_A)+(N\!-\!2)(N\!-\!3)n(\bc_B)}=\frac{2}{(N\!-\!2)(2N\!-\!5)}
\end{eqnarray}
Hence, if we measure during the `accept all' edge swap process the relative frequency graph distribution
\begin{equation}
Q(f)=\frac{1}{|\Omega[\bk]|}\sum_{\bc\in\Omega[\bk]}\delta\Big[f-|\Omega[\bk]|p_\infty(\bc)\Big]
\end{equation}
we should find
\begin{equation}
Q(f)=\frac{1}{1\!+\!(N\!-\!2)(N\!-\!3)}
\delta\Big[f-\frac{1\!+\!(N\!-\!2)(N\!-\!3)}{2N\!-\!5}\Big]
+\frac{(N\!-\!2)(N\!-\!3)}{1\!+\!(N\!-\!2)(N\!-\!3)}
\delta\Big[f-\frac{2[1\!+\!(N\!-\!2)(N\!-\!3)]}{(N\!-\!2)(2N\!-\!5)}\Big]
\end{equation}
(as opposed to the result $Q(f)=\delta(f-1)$ that would be obtained for a flat measure on $\Omega[\bk]$).
We have carried out numerical simulations of the edge swap dynamics (\ref{eq:edgeswapMC1},\ref{eq:edgeswapMC2}), first on an `accept all moves' basis and second using the canonical acceptance probabilities $A(\bc|\bc^\prime)=[1+n(\bc)/n(\bc^\prime)]^{-1}$, for graphs as in Figure \ref{milo2} with $N=10$. Here one expects to find $Q(f)=\delta(f-1)$ for canonical acceptance probabilities, but for the `accept all' edge swap dynamics we should get:
\begin{eqnarray}
N=10:&~~~~&
Q(f)=\frac{1}{57}
\delta\Big[f-\frac{19}{5}\Big]
+\frac{56}{57}
\delta\Big[f-\frac{57}{60}\Big]
\end{eqnarray}
The results are shown in figure \ref{milo3}. The distribution $p_\infty(\bc)$ generated by the `accept all' edge swap dynamics is indeed not uniform: the smaller peak in $Q(f)$ appearing at the predicted value $f=19/5=3.8$ reflects the visits to the type A graph, which occur more frequently due to its larger mobility.
Conversely,
 with the canonical acceptance probability $A(\bc|\bc^\prime)=n(\bc^\prime)/[n(\bc)+n(\bc^\prime)]$ (which here takes the three possible values
 $A(\bc_A|\bc_B)=0.2$, $A(\bc_B|\bc_B)=0.5$, and $A(\bc_B|\bc_A)=0.8$) the resulting equilibrium  measure $p_\infty(\bc)$ of the process is indeed flat, i.e. $Q(f)=\delta(f-1)$.
\vsp

\begin{figure}[t]
\vspace*{0mm} \hspace*{8mm} \setlength{\unitlength}{0.68mm}
\begin{picture}(200,90)

  \put(0,0){\includegraphics[height=90\unitlength]{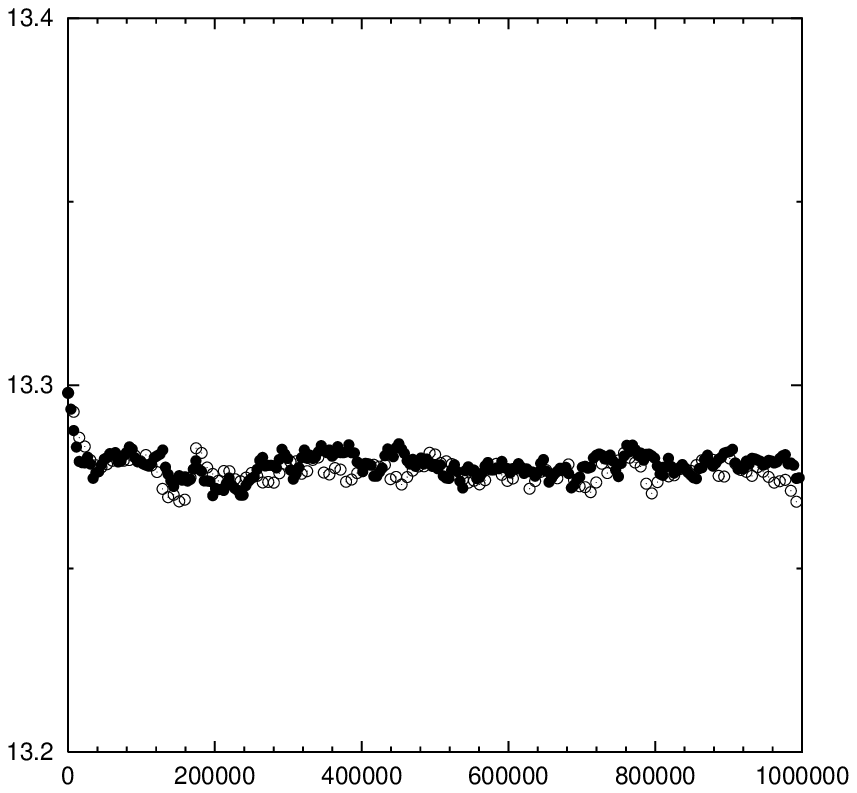}}
 \put(-15,50){$\overline{n(\bc)}/N^2$}
 \put(35,-9){\em executed moves}

  \put(115,0){\includegraphics[height=90\unitlength]{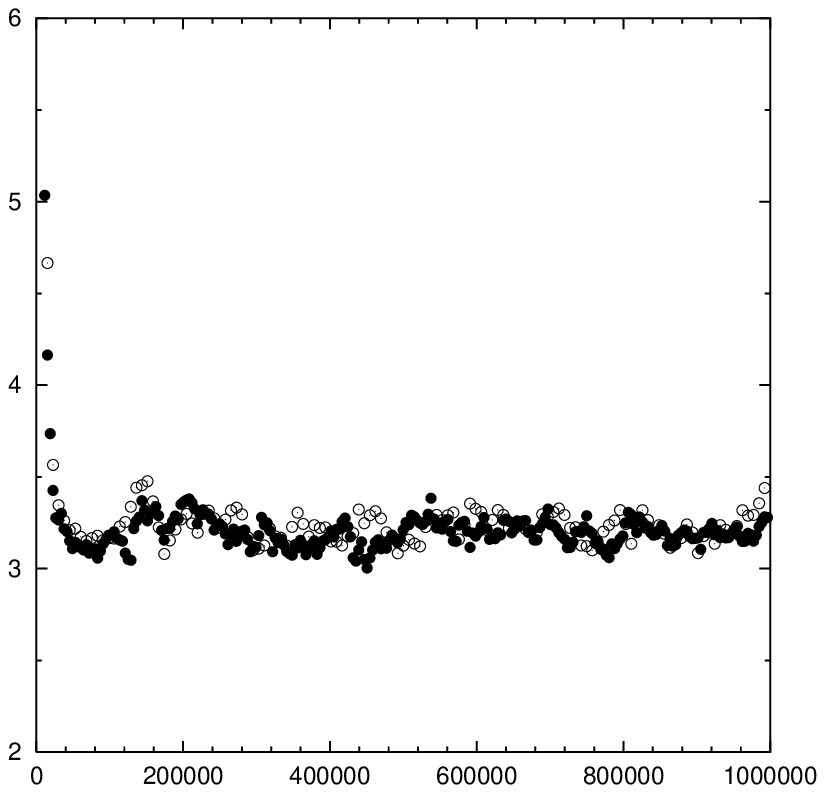}}
 \put(98,50){${\rm Tr}(\bc^3)/N$}
 \put(150,-9){\em executed moves}

\end{picture}
 \vspace*{10mm}
\caption{Left: re-scaled running averages $\overline{n(\bc)}/N^2$ of the graph mobility (\ref{eq:nc}), measured during edge swap dynamics of the type (\ref{eq:edgeswapMC1},\ref{eq:edgeswapMC2}) in the human protein interaction network \cite{hprd} with $N=9463$ nodes and average connectivity $\bra k\ket\approx 7.402$. The time unit is the number of executed edge swaps, and running averages are
measured over time windows of about $3800$ successive states $\bc$. Full circles: `accept all' edge swap dynamics, i.e.
 $A(\bc|\bc^\prime)=1$. Open circles: randomly proposed and possible edge swaps are executed with the canonical acceptance rates (\ref{eq:acceptance_flat}), i.e. $A(\bc|\bc^\prime)=[1+n(\bc)/n(\bc^\prime)]^{-1}$.
Right: corresponding measurements of ${\rm Tr}(\bc^3)/N$, which is proportional to the average number of length-3 loops per node in the network.
}
\label{fig:human_hprd}
\end{figure}

In order to assess to what extent the differences between `accept all' and correct edge swap dynamics manifest themselves in large and realistic graphs as studied intensively in biology and bio-informatics, we have also carried out edge swap simulations (similar to those described above for synthetic graphs) in protein interaction networks. We chose the most recent data for the human protein interaction network \cite{hprd}, giving a graph of $N=9463$ nodes and degree statistics $\bra k\ket\approx 7.402$ and $\bra k^2\ket\approx 248.7$ (with $k_{\rm max}=247$).
For this data set the simple condition (\ref{eq:conditions_flat_measure}) does not yet permit us to rely on `accept all' edge swap randomization as a safe algorithm for unbiased randomization, since here one finds
\begin{eqnarray}
\bra k^2\ket k_{\rm max}/\bra k\ket^2  N\approx 0.12
\end{eqnarray}
As with the synthetic graphs, we ran
two different Monte-Carlo processes as described by (\ref{eq:edgeswapMC1},\ref{eq:edgeswapMC2}): `accepted all' edge swaps dynamics, i.e. $A(\bc|\bc^\prime)=1$, and edge swap dynamics with acceptance rates (\ref{eq:acceptance_flat}), i.e.  $A(\bc|\bc^\prime)=[1+n(\bc)/n(\bc^\prime)]^{-1}$ (to guarantee a uniform equilibrium measure). The results are shown in figure \ref{fig:human_hprd}. Here the differences between the two types of dynamics are seen to be negligible, both in terms of the observed mobility $n(\bc)$ and in terms of quantities such as ${\rm Tr}(\bc^3)$ (which is equivalent to counting the number of length-3 loops in the network). One finds similar results for the available protein interaction data of other organisms.  This is a relevant observation, since graph randomization via `accept all' edge swapping has in the past been used to quantify the relative frequency of small network modules of `motifs' in biological networks, of which length-3 loops are just a primitive example, or to quantify the relevance of observed degree-degree correlations; see e.g. \cite{MaslovSneppen,Batada,Friedel}.  It follows that those who have in the past used `accept all' edge swap dynamics to randomize protein interaction networks have been fortunate, in that for the available data the incorrect sampling resulting from this dynamics does not appear to interfere with observation. However, since it is now generally agreed that the presently available incomplete protein interaction data are {\em biased} samples  of the full proteome, there is no guarantee that when data have become more complete and representative the simple but in principle incorrect `accept all' edge swap randomization will continue to work in practice.

\subsection{Simulations to produce controlled non-uniform measures}

\begin{figure}[t]
\vspace*{5mm} \hspace*{8mm} \setlength{\unitlength}{0.64mm}
\begin{picture}(200,210)

  \put(5,120){\includegraphics[height=85\unitlength]{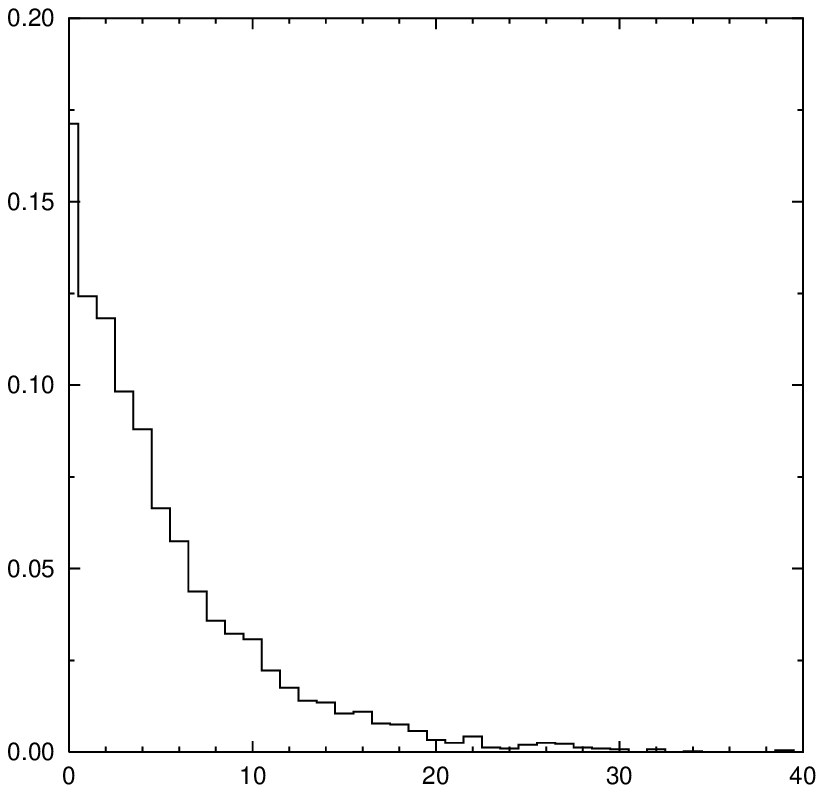}}
 \put(-5,166){$P(k)$}
 \put(53,111){$k$}

  \put(115,216){\includegraphics[height=110\unitlength,angle=270]{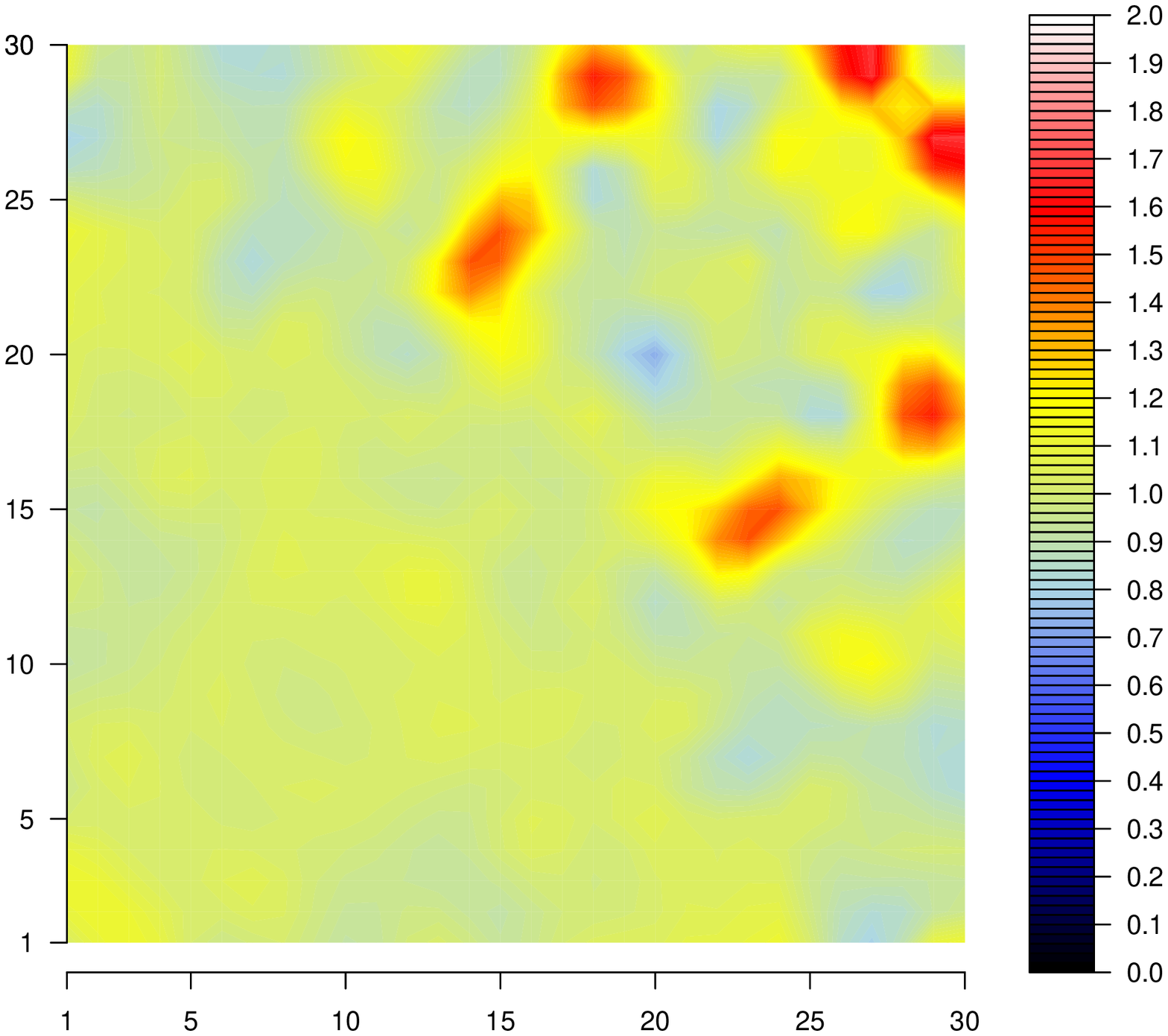}}
 \put(115,166){$k^\prime$}
 \put(163,111){$k$}
 \put(155,206){$\Pi(k,k^\prime|\bc_0)$}

 \put(0,110){\includegraphics[height=110\unitlength,angle=270]{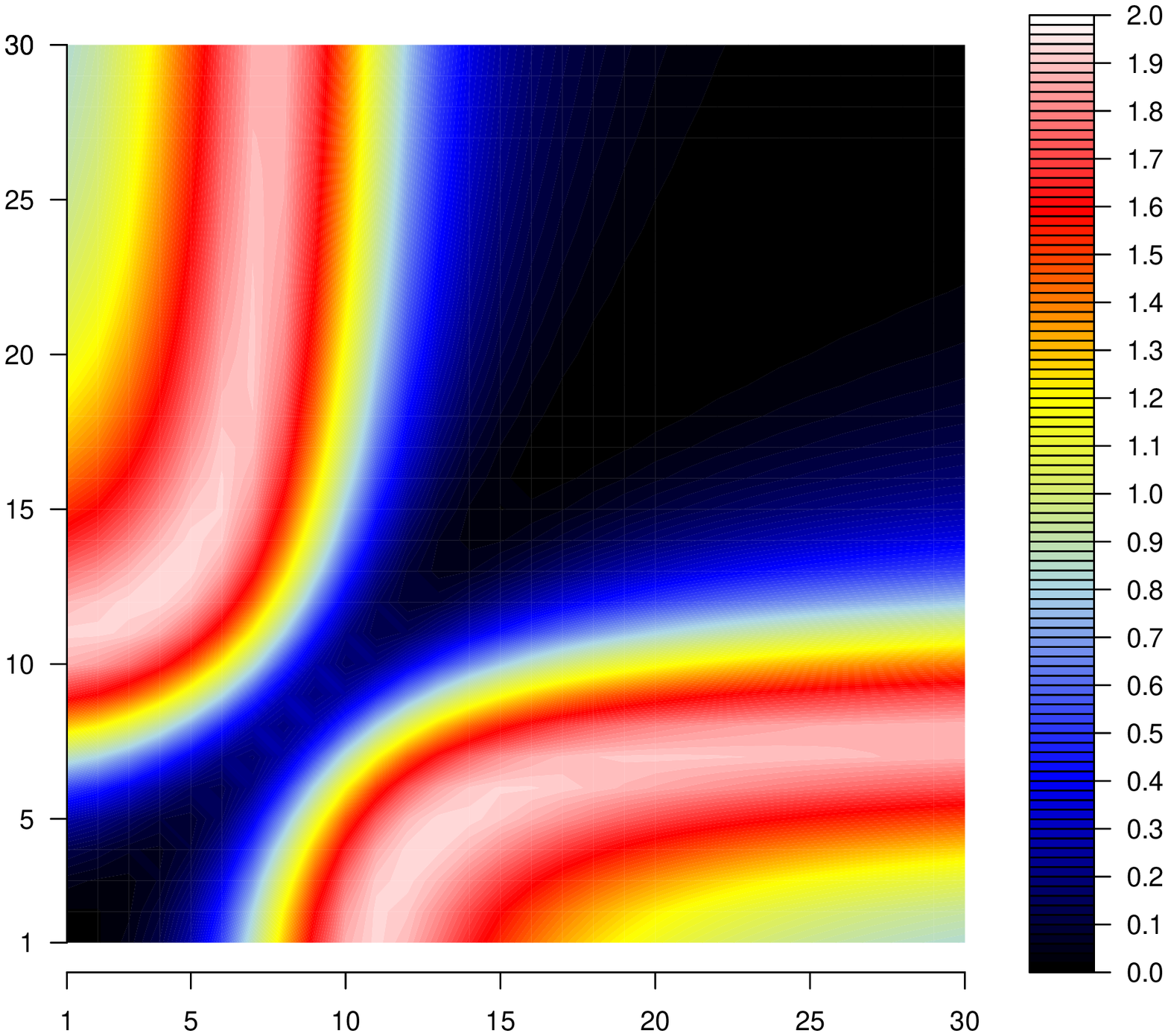}}
 \put(-2,60){$k^\prime$}
 \put(48,5){$k$}
 \put(35,100){$\Pi(k,k^\prime)$ (theory)}

  \put(115,110){\includegraphics[height=110\unitlength,angle=270]{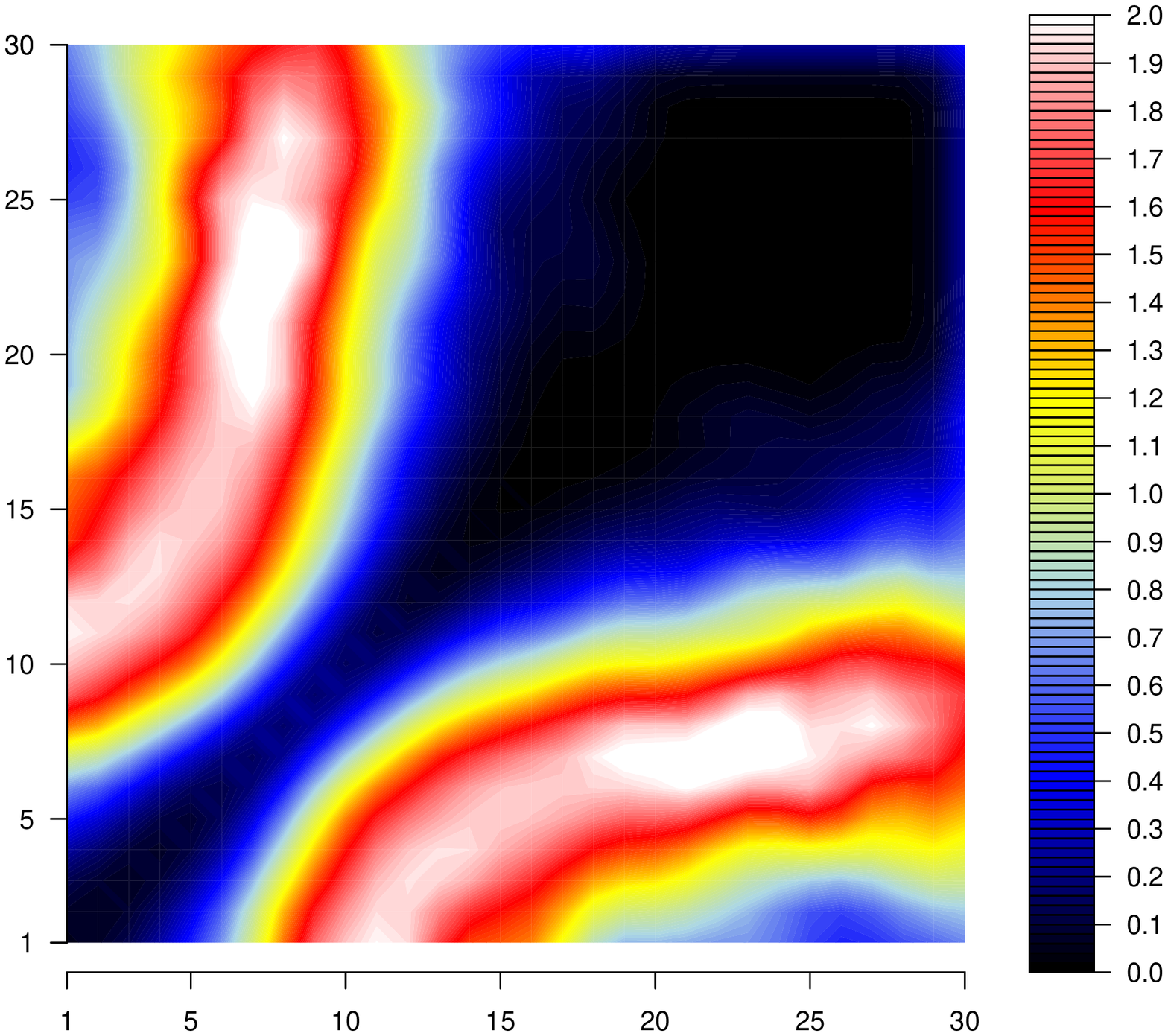}}
 \put(115,60){$k^\prime$}
 \put(163,5){$k$}
 \put(155,100){$\Pi(k,k^\prime|\bc_{\rm final})$}

\end{picture}
 \vspace*{0mm}
\caption{Results of edge-swap simulations tailored to generating equilibrium random graph ensembles with the non-uniform measure
(\ref{eq:newer_connectivity}). Top left: degree distribution of the graph, with $N=4000$ and $\bra k\ket=5$.
Top right: colour plot of the relative degree correlations $\Pi(k,k^\prime|\bc_0)$ as measured in the (randomly generated) initial graph $\bc_0$.
Bottom left: colour plot of the predicted relative degree correlations (\ref{eq:Pi_predicted}), corresponding to $Q(k,k^\prime)=(k\!-\!k^\prime)^2/C$, which is found for $N\to\infty$ in the ensemble (\ref{eq:newer_connectivity}) (the target of our edge swap process).
Bottom right: colour plot of the relative degree correlations $\Pi(k,k^\prime|\bc_{\rm final})$ in the final graph $\bc_{\rm final}$, measured after 75,\!000 accepted moves of the
Markov chain
(\ref{eq:edgeswapMC1},\ref{eq:edgeswapMC2},\ref{eq:flipenergies}) with canonical acceptance rates (\ref{eq:PC_A},\ref{eq:PC_deltaH},\ref{eq:PC_L}). The latter is indeed in good agreement with the prediction (bottom left) corresponding to the target measure (\ref{eq:newer_connectivity}).}
\label{fig:Pi}
\end{figure}

Our final simulations involved Markov chains of the form (\ref{eq:edgeswapMC1},\ref{eq:edgeswapMC2},\ref{eq:flipenergies}) tailored to evolve towards controlled non-uniform equilibrium measures. Here we tested the prediction that the edge swap process with edge swap acceptance rates defined by  (\ref{eq:PC_A},\ref{eq:PC_deltaH},\ref{eq:PC_L})
will evolve towards the nontrivial measure (\ref{eq:newer_connectivity}).
To test this, we measured the relative degree correlations $\Pi(k,k^\prime|\bc)$ as defined in (\ref{eq:Pi}) upon equilibrating the edge swap dynamics
(\ref{eq:edgeswapMC1},\ref{eq:edgeswapMC2},\ref{eq:flipenergies}) with acceptance rates (\ref{eq:PC_A},\ref{eq:PC_deltaH},\ref{eq:PC_L}), and compared the result with the theoretical prediction extracted from (\ref{eq:newer_connectivity}).
For the kernel $Q(k,k^\prime)$ in (\ref{eq:newer_connectivity}) we took   $Q(k,k^\prime)=C^{-1}(k-k^\prime)^2$, with normalization dictating that $C=2(\bra k^2\ket-\bra k\ket^2)$. For sufficiently large $N$, and given that the measure is indeed (\ref{eq:newer_connectivity}), the predicted values for $\Pi(k,k^\prime)$ are
\begin{eqnarray}
\Pi(k,k^\prime)&=& C^{-1}(k-k^\prime)^2/[F(k)F(k^\prime)]
\end{eqnarray}
where $F(k)$ is to be solved from (\ref{eq:solvingF}), which here gives
$F(k)=(\alpha_3-2\alpha_2 k+\alpha_1 k^2)/\sqrt{\bra k\ket C}$,
in which the three coefficients $\alpha_\ell$ are to be solved numerically from
\begin{eqnarray}
\alpha_\ell&=&\sum_k  \frac{k^{\ell} p(k)}{\alpha_3-2\alpha_2 k+\alpha_1 k^2}
\label{eq:alphas}
\end{eqnarray}
The predicted relative degree correlations are then given by
\begin{eqnarray}
\Pi(k,k^\prime)&=& \frac{\bra k\ket (k-k^\prime)^2}{[\alpha_3-2\alpha_2 k+\alpha_1 k^2][\alpha_3-2\alpha_2 k^\prime+\alpha_1 k^{\prime 2}]}
\label{eq:Pi_predicted}
\end{eqnarray}
We generated a simple synthetic initial graph $\bc_0$ with $N=4000$ and $\bra k\ket=5$, with the non-Poissonian degree distribution shown in figure \ref{fig:Pi} (top left). Its  relative degree correlations $\Pi(k,k^\prime|\bc_0)$ were found to be all close to one (being the value for all $\Pi(k,k^\prime)$ that one would have found for the flat ensemble $p(\bc)=Z^{-1}\prod_i \delta_{k_i,k_i(\bc)}$); see figure \ref{fig:Pi} top right. After iterating the Markov chain
(\ref{eq:edgeswapMC1},\ref{eq:edgeswapMC2},\ref{eq:flipenergies}), with canonical acceptance rates (\ref{eq:PC_A},\ref{eq:PC_deltaH},\ref{eq:PC_L}) tailored to approaching (\ref{eq:newer_connectivity}) as an equilibrium measure, and upon reaching what appeared to be equilibration (after 75,\!000 accepted edge swaps), we found indeed values for the degree correlations that were in very good agreement with those corresponding to
the nontrivial target ensemble (\ref{eq:newer_connectivity}) (shown in the bottom panels of figure \ref{fig:Pi}). Note that perfect agreement is expected only for $N\to\infty$.

%%%%%%%%%%%%%%%%%%

\section{Conclusions}

The mathematical analysis of stochastic processes occurring in the space of graphs with prescribed properties presents a number of intriguing and challenging features, that have just started to be addressed in the language of statistical mechanics. From a physical viewpoint, understanding graph dynamics is important in particular when seeking to discern the basic features of graphs (the constraints) from the ones that are induced by them, which generically characterize the ensemble of graphs satisfying the constraints. It turns out that the mobility, namely the number of allowed moves away from a given state, is a central quantity for unraveling several properties specific of such processes. In particular, the very dependence of the mobility on the specific state (which in principle can persist even in an infinite graph) is responsible for important entropic effects that may prevent a simple switching dynamics (simple in the sense that all randomly generated and possible moves are executed) from sampling uniformly the space of graphs compatible with given constraints. Uniform sampling is especially desirable in applications where one is either interested in testing the robustness of certain graphical properties against graph ``randomization'' or where one aims to generate random graphs that satisfy a given set of constraints but are otherwise unbiased.

In this paper we have constructed a general framework for constrained stochastic graph dynamics, and derived an explicit and relatively simple expression for the graph mobility in the case where the dynamics is based on randomly generated `edge swaps' or `switchings'.
This latter expression allowed us to: (i) define Monte-Carlo processes that are guaranteed to converge to any desired measure on the space of graphs with a prescribed degree sequence, (ii) calculate explicitly the equilibrium measure that would be found for `accept all' edge swap dynamics (which will generally not be uniform), and
(iii) identify practical conditions on the graph topology that are sufficient to ensure that even the equilibrated simple `accept all' edge swap dynamics will give a uniform dynamical sampling of the accessible graphs in the limit where the number of nodes $N\to\infty$.

  We have carried out Monte-Carlo simulations of both synthetic graphs and of biological networks describing protein-protein interactions, in which we compared the results of executing `accept all' edge swap dynamics versus edge swap dynamics with correct acceptance probabilities tailored to producing unbiased equilibrium measures. We also carried out simulations of stochastic edge swap dynamics that are tailored to generating random graph ensembles with controlled non-uniform measures, characterized by nontrivial imposed degree correlations.
 All numerical simulations lend convincing support to our theoretical predictions, and underline the importance of the graph mobility in the construction of correct Monte-Carlo acceptance probabilities in constrained graph dynamics.

We have taken the approach of formulating stochastic graph dynamics within a general statistical mechanics framework, where the constrained Markov dynamics of graphs is treated similarly to a spin-flip dynamics in magnetic systems (albeit that in the latter the issue of state mobility does not arise, not even in the case of constrained Kawasaki-type dynamics \cite{kawa}). This suggests that future studies of graph dynamics may well reveal a rich (and possibly unexpected) phenomenology paralleling that of magnetic spin systems.

%%%%%%%%%%%%%%%%%

\begin{acknowledgements}
It is our pleasure to thank M. Marsili and F. Ricci Tersenghi for thoughtful comments and suggestions, and L. Fernandes for kindly providing us with protein interaction network data.
\end{acknowledgements}

%%%%%%%%%%%%%%%%%%%%%

%%%%%%%%%%%%%%%%%%%%%%

\clearpage
\appendix

\section{Edge swaps as minimal degree-preserving moves in graphs}
\label{app:switches}

For the benefit of the reader, we review in this appendix briefly in the language of the present study the arguments leading to the identification of edge swaps as the minimal degree preserving moves.
Let us first characterize
 all possible moves $F:\bc\to\bc^\prime\neq\bc$ that leave the degrees $k_i(\bc)=\sum_j c_{ij}$ of all $N$ nodes in a graph invariant. Since $c_{ij}\in\{0,1\}$ for all $(i,j)$, and $c_{ij}=c_{ji}$,
each move $F$ is characterized fully by specifying the set $S_F=\{(i,j)|~i<j,~Fc_{ij}=1-c_{ij}\}$ of node pairs that are affected by $F$.
If all degrees are conserved, then also the average connectivity is conserved, so each link removed by $F$ must be compensated elsewhere in the graph by a link created. Hence  $|S_{F}|$ is even.
We define $[a,b]=(a,b)$ if $a<b$ and $[a,b]=(b,a)$ if $b<a$, and $S_{F,i}=\{ j | [i,j]\in S_F \}$ (i.e. the set of
those nodes $j$ that share with $i$ a
link that is either removed or created by the move $F$).
 Our first question is then: for which sets $S_F$ of node pairs will the collective inversion of all links $c_{ij}\to 1-c_{ij}$ preserve all degrees, i.e. obey
\begin{eqnarray}
{\rm for~all~}i:&~~~&0= \frac{1}{2|S_{F,i}|}\sum_{j\in S_{F,i}}[c_{ij}-(1-c_{ij})] =\frac{1}{|S_{F,i}|}\sum_{j\in S_{F,i}}c_{ij}-\frac{1}{2}
\label{eq:conditions_F}
\end{eqnarray}
For each node $i$, there is an even number of pairs $(i,j)$ affected by $F$, of which half see a link removed ($c_{ij}=1\to Fc_{ij}=0$) and half see a link created ($c_{ij}=0\to Fc_{ij}=1$). Hence also $|S_{F,i}|$ must be even for each $i$ with bonds created or removed by $F$. Since It is clear that whether or not a set $S_F$ (i.e. a move $F$) meets the $N$ conditions (\ref{eq:conditions_F}) {\em must} depend on the graph $\bc$ at hand.

The {\em minimal} moves that satisfy (\ref{eq:conditions_F}) are defined as those involving the smallest set $S_F$. Let $i$ be a site with bonds created or removed by $F$. Since $|S_{F,i}|$ is even, there will be at least two further nodes $(j,k)$ with bonds created or removed by $F$. Each of these must have their own sets $S_{F,j}$ and $S_{F,k}$ of even size, so it impossible for the the action of $F$ to be restricted to the node pairs $[i,j]$ and $[i,k]$ alone. Thus the smallest possible size for $S_F$ (which we know to be even) is $|S_F|=4$. Let us inspect the properties of moves with $|S_F|=4$  in detail:
\vsp

\begin{itemize}
\item[(i)]
Each node $i$ involved in $F$ has
$|S_{F,i}|=2$, i.e. participates in precisely two of the four node pairs in $S_F$.
\\[2mm]
Proof:\\
According to (\ref{eq:conditions_F}) each such $i$ participates in at least two of the four node pairs in $S_F$. If we had
$|S_{F,i}|>2$, then there would be at least four other nodes involved in $F$ (since $|S_{F,i}|$ is even), each in turn participating in at least two pairs in $S_F$.
The minimal set $S_{F}$ would then contain $S_F=\{[i,j],[i,k],[i,\ell],[i,m],[k,\star],\ldots\}$ (where $\star$ is some node not equal to $i$ or $k$), which contradicts $|S_F|=4$.
\vsp

\item[(ii)]
$F$ involves exactly four distinct nodes $(i,j,k,\ell)$, and $S_F=\{[i,j],[i,k],[j,\ell],[k,\ell]\}$.
\\[2mm]
Proof:\\
Starting from any involved node $i$, with $S_{F,i}=\{j,k\}$, we know that $S_F$ contains $S_F=\{[i,j],[i,k],[k,\star],\ldots\}$, where $\star$ is a node such that $S_{F,k}=\{i,\star\}$ (so $\star\notin\{i,k\}$). The first possibility is $\star=j$. But if this were the case then
either $S_{F}=\{[i,j],[i,k],[k,j]\}$ (contradicting $|S_F=4|$), or  $S_{F}=\{[i,j],[i,k],[k,j],[\ell,m],\ldots\}$ for some new indices $(\ell,m)$. In the latter case, however, since $|S_{F,\ell}|=|S_{F,m}|=2$, the set $S_F$ must contain additional node pairs involving $\ell$ and $m$, giving $|S_{F}|>4$. It follows that $\star=j$ is not allowed. This leaves us with $S_F=\{[i,j],[i,k],[k,\ell],[j,\ell]\}$ for some $\ell\notin\{i,j,k\}$ as the only option.
\vsp
\end{itemize}

\noindent
Since we know that $F$ acts as $c_{ij}\to 1- c_{ij}$ for all $(i,j)\in S_F$, and that before the move exactly half of the pairs $(i,j)\in S_F$ have $c_{ij}=1$, the minimal moves are (modulo node permutations) of the form of so-called `edge swaps' as shown below (where thick lines indicate $c_{ab}=1$ and thin lines indicate $c_{ab}=0$):

\setlength{\unitlength}{0.2mm}
 \begin{picture}(500,170)(-170,0)
 \put(5,0){\here{\bcirc}}\put(105,0){\here{\bcirc}}\put(5,100){\here{\bcirc}}\put(105,100){\here{\bcirc}}
 \put(0,125){\here{$i$}}\put(100,125){\here{$j$}}\put(100,-25){\here{$k$}}\put(0,-25){\here{$\ell$}}
 \thinlines \put(0,0){\line(0,1){100}} \put(100,0){\line(0,1){100}}
 \thicklines\put(0,100){\line(1,0){100}}\put(0,101){\line(1,0){100}}\put(0,99){\line(1,0){100}}
 \put(0,0){\line(1,0){100}} \put(0,-1){\line(1,0){100}}\put(0,1){\line(1,0){100}}

\put(200,50){\large\here{$\Longrightarrow$}} \put(200,85){\large\here{$F$}}

\put(305,0){\here{\bcirc}}\put(405,0){\here{\bcirc}}\put(305,100){\here{\bcirc}}\put(405,100){\here{\bcirc}}
 \put(300,125){\here{$i$}}\put(400,125){\here{$j$}}\put(400,-25){\here{$k$}}\put(300,-25){\here{$\ell$}}

 \thicklines
 \put(300,0){\line(0,1){100}} \put(299,0){\line(0,1){100}} \put(301,0){\line(0,1){100}}
 \put(400,0){\line(0,1){100}} \put(399,0){\line(0,1){100}}\put(401,0){\line(0,1){100}}
 \thinlines\put(300,100){\line(1,0){100}} \put(300,0){\line(1,0){100}}

\end{picture}
\vspace*{10mm}

 \noindent
 It is clear that these transitions preserve all degrees of a graph. We now know also that
 these are the {\em simplest} nontrivial transitions with this property. It is fairly straightforward to generalize the above
 representation, and show that each allowed move $F$ with $|S_F|=m$ corresponds to a set $S_F$ that describes a closed path $(i_1\to i_2\to \ldots i_m\to i_1)$ connecting $m$ nodes, such that all $c_{i_{\ell-1} i_{\ell}}=1-c_{i_{\ell} i_{\ell+1}}$ (with $\ell~{\rm mod}~m$). A path is allowed to cross itself, with $\frac{1}{2}|S_{F,i}|-1$ giving the number of crossings at node $i$, but is not allowed to have overlapping segments.
The action of $F$ is then the inversion $c_{ij}\to 1-c_{ij}$ of all bond variables along the path.

\clearpage
\section{Upper and lower bounds for the graph mobility $n(\bc)$}
\label{app:nc_bounds}

 Here we establish simple bounds on the quantity $n(\bc)$ defined in (\ref{eq:nc}), expressed solely in terms of the degree moments $\bra k\ket$ and $\bra k^2\ket$ and the maximum degree $k_{\rm max}=\max_i k_i$.
First we inspect the term with ${\rm Tr}(\bc^3)$. The only possible general lower bound is the trivial ${\rm Tr}(\bc^3)\geq 0$, since this is satisfied by all tree-like graphs with arbitrary degree distributions. To obtain an upper bound we use the inequality $c_{jk}c_{ki}\leq \frac{1}{2}[c_{jk}+c_{ki}]$:
\begin{eqnarray}
{\rm Tr}(\bc^3)&=& \sum_{ijk}c_{ij}c_{jk}c_{ki}\leq \frac{1}{2}\sum_{ijk}c_{ij}[c_{jk}+c_{ki}]
= \sum_{i}k_i^2=N\bra k^2\ket
\label{eq:Tr3_upperbound}
\end{eqnarray}
Next we turn to ${\rm Tr}(\bc^4)$. A suitable lower bound can be constructed as follows:
\begin{eqnarray}
{\rm Tr}(\bc^4)&=& \sum_{ijk\ell}c_{ij}c_{jk}c_{k\ell}c_{\ell i}\geq \sum_{ijk\ell}c_{ij}c_{jk}c_{k\ell}c_{\ell i}\delta_{j\ell}
=\sum_{ijk}c_{ij}c_{jk}=\sum_j k_j^2=N\bra k^2\ket
\label{eq:Tr4_lowerbound}
\end{eqnarray}
An upper bound follows from  $c_{ij}c_{k\ell}\leq \frac{1}{2}(c_{ij}+c_{k\ell})$:
\begin{eqnarray}
{\rm Tr}(\bc^4)&\leq & \frac{1}{2}\sum_{ijk\ell}[c_{ij}+c_{k\ell}]c_{jk}c_{\ell i}
=\sum_{ij}k_ic_{ij}k_{j}
\label{eq:Tr4_upperbound}
\end{eqnarray}
It follows from
the four bounds constructed so far
that
\begin{eqnarray}
\frac{1}{4}N\bra k^2\ket\leq
\frac{1}{4}{\rm Tr}(\bc^4)+\frac{1}{2}{\rm Tr}(\bc^3)
\leq \frac{1}{2}N\bra k^2\ket +\frac{1}{4}\sum_{ij}k_ic_{ij}k_{j}
\end{eqnarray}
Inserting into
formula (\ref{eq:nc}) subsequently gives
\begin{eqnarray}
n(\bc)&\geq &\frac{1}{4}N^2\bra k\ket^2 +\frac{1}{4}N\bra k\ket
-\frac{1}{4}N\bra k^2\ket
-\frac{1}{2}\sum_{ij}k_{i}c_{ij}k_j
\\
n(\bc)&\leq &
\frac{1}{4}N^2\bra k\ket^2 +\frac{1}{4}N\bra k\ket
-\frac{1}{4}\sum_{ij}k_{i}c_{ij}k_j
\end{eqnarray}
To proceed we need bounds for the term $\sum_{ij}k_{i}c_{ij}k_j$. A simple lower bound follows from the fact that if $c_{ij}=1$ then $k_i\geq 1$.
An upper bound follows from $k_i\leq k_{\rm max}$, and so we get
\begin{eqnarray}
\sum_{ij}k_{i}c_{ij}k_j&\geq& \sum_{ij}c_{ij}k_j=\sum_j k_j^2=N\bra k^2\ket
\\
\sum_{ij}k_{i}c_{ij}k_j&\leq& k_{\rm max}\sum_{ij}c_{ij}k_j=Nk_{\rm max}\bra k^2\ket
\end{eqnarray}
This then leads to the following remarkably tight bounds for the mobility (note that always $\bra k^2\ket\geq \bra k\ket$):
\begin{eqnarray}
n(\bc)&\geq &\frac{1}{4}N^2\bra k\ket^2 +\frac{1}{4}N\bra k\ket
-\frac{1}{4}N\bra k^2\ket( 2k_{\rm max}+1)
\label{eq:lower_bound_nc}
\\
n(\bc)&\leq &
\frac{1}{4}N^2\bra k\ket^2 +\frac{1}{4}N\bra k\ket
-\frac{1}{4}N\bra k^2\ket
\label{eq:upper_bound_nc}
\end{eqnarray}
A further corollary from this is an absolute bound on the mobility change $\Delta_{ijk\ell;\alpha}n(\bc)=F_{ijk\ell;\alpha}n(\bc)-n(\bc)$ due to a single
edge swap:
\begin{eqnarray}
|\Delta_{ijk\ell;\alpha}n(\bc)|&\leq &
\frac{1}{2}N\bra k^2\ket k_{\rm max}
\label{eq:swap_bound_nc}
\end{eqnarray}

\clearpage

\section{Effect of single edge swaps on ${\rm Tr}(\bc^3)$ and ${\rm Tr}(\bc^4)$}
\label{app:traces}

Here we study the terms in (\ref{eq:nc}) that involve traces. We limit ourselves to states $\bc$ on which the edge swap operator $F_{ijkl;\alpha}$ can act, since only those are required in (\ref{eq:flipenergies}).
First we introduce some further notation. We define $[a,b]=(a,b)$ if $a<b$ and $[a,b]=(b,a)$ if $b<a$, and we denote the
relevant sets of index pairs as follows:
\begin{eqnarray}
&S_{pqv}=\{[p,q],[q,v],[v,p]\},~~~~~~       &S_{pqv;ijk\ell;\alpha}=S_{pqv}\cap S_{ijk\ell;\alpha}\\
&S_{pqvw}=\{[p,q],[q,v],[v,w],[w,p]\},~~~~~~&S_{pqvw;ijk\ell;\alpha}=S_{pqvw}\cap S_{ijk\ell;\alpha}
\end{eqnarray}
with in the first line $p\neq q$, $q\neq v$ and $v\neq p$, and in the second
line $p\neq q$, $q\neq v$, $v\neq w$, and $w\neq p$.

Also, we recall that always $i<j<k<\ell$,
and we associate to every set of index pairs $S_{ijk\ell;\alpha}$
a closed path ${\cal P}^{[a,b]}_{ijkl;\alpha}$ with $[a,b]\in S_{ijkl}$,
starting from $a$, along the
$4$ possible bonds through each pair of indices in the set $S_{ijkl;\alpha}$, and passing through the link $[a,b]$ in the order  $a$ to $b$.
Index pairs with one index in common are
visited sequentially. Hence,
{\it e.g.} ${\cal P}^{[i,j]}_{ijkl;1}$ is uniquely
determined as the closed path $i\rightarrow j\rightarrow k\rightarrow
\ell\rightarrow i$.
Finally, we indicate by $b^{[a,b]}_{+;\alpha}$ and
$a^{[a,b]}_{-;\alpha}$ the index that
follows $b$, and the index that preceeds $a$, respectively,
along the closed path
${\cal P}^{[a,b]}_{ijkl;\alpha}$. For instance,
 $j^{[i,j]}_{+;1}=k$ and $i^{[i,j]}_{-;1}=\ell$, by periodicity
(note that by definition $b=a^{[a,b]}_{+;\alpha}$ and $a=b^{[a,b]}_{-;\alpha}$).

\subsection{Full expressions for $\Delta_{ijk\ell;\alpha}{\rm Tr}(\bc^3)$ and $\Delta_{ijk\ell;\alpha}{\rm Tr}(\bc^4)$}

We recall that the edge swap operator $F_{ijk\ell;\alpha}$ can only affect the presence or absence of bonds in the set ${\cal S}_{ijk\ell;\alpha}$ (where $|{\cal S}_{ijk\ell;\alpha}|= 4$),
and that its action is always to create two new bonds and destroy two present ones. Hence
\begin{eqnarray}
\Delta_{ijk\ell;\alpha}{\rm Tr}(\bc^3)&=& \sum_{n=1}^3 \sum_{pqv}\delta_{|S_{pqv;ijk\ell;\alpha}|,n}
~\Delta_{ijk\ell;\alpha}(c_{[p,q]}c_{[q,v]}c_{[v,p]})
\\
\Delta_{ijk\ell;\alpha}{\rm Tr}(\bc^4)&=& \sum_{n=1}^4 \sum_{pqvw}\delta_{|S_{pqvw;ijk\ell;\alpha}|,n}
~\Delta_{ijk\ell;\alpha}(c_{[p,q]}c_{[q,v]}c_{[v,w]}c_{[w,p]})
\end{eqnarray}
In fact, it turns out that all terms with $n>1$ must be zero:
\begin{itemize}
\item
For each $n=2$ term there are two index pairs  $(x,y)\in\{[p,q],[q,v],[v,p]\}$ in the case of ${\rm Tr}(\bc^3)$ and two index pairs  $(x,y)\in\{[p,q],[q,v],[v,w],[w,p]\}$ in the case of ${\rm Tr}(\bc^4)$ such that $F_{ijkl;\alpha}c_{xy}=1-c_{xy}$. If both pairs have $c_{xy}=1$, or both pairs have $c_{xy}=0$, then the conditions (\ref{eq:Fcond1},\ref{eq:Fcond2},\ref{eq:Fcond3}) for $F_{ijk\ell;\alpha}$ to act dictate
that these two pairs have no indices in common.
So the options for the action of $F_{ijkl;\alpha}$ are (modulo permutations):
\begin{eqnarray*}
{\rm Tr}(\bc^3):&~~&\{c_{[p,q]},c_{[q,v]},c_{[v,p]}\}=\{1,1,0\}~\to~\{1,0,1\}\\
                &~~&\{c_{[p,q]},c_{[q,v]},c_{[v,p]}\}=\{1,0,0\}~\to~\{0,1,0\}
                \\[2mm]
{\rm Tr}(\bc^4):
&~~&\{c_{[p,q]},c_{[q,v]},c_{[v,w]},c_{[w,p]}\}=\{1,1,1,1\}~\to~\{1,0,1,0\}\\
&~~&\{c_{[p,q]},c_{[q,v]},c_{[v,w]},c_{[w,p]}\}=\{1,1,1,0\}~\to~\{0,1,0,0\}~{\rm or}~\{1,1,0,1\}\\
&~~&\{c_{[p,q]},c_{[q,v]},c_{[v,w]},c_{[w,p]}\}=\{1,0,1,0\}~\to~\{0,0,0,0\}~{\rm or}~\{1,1,0,0\}~{\rm or}~\{1,1,1,1\}\\
&~~&\{c_{[p,q]},c_{[q,v]},c_{[v,w]},c_{[w,p]}\}=\{1,0,0,0\}~\to~\{1,1,0,1\}~{\rm or}~\{0,1,0,0\}\\
&~~&\{c_{[p,q]},c_{[q,v]},c_{[v,w]},c_{[w,p]}\}=\{0,0,0,0\}~\to~\{1,0,1,0\}
\end{eqnarray*}
All but two cases have at least one $c_{xy}=0$ before the swap and at least one $c_{xy}=0$ after the swap; for those we know immediately
that $\Delta_{ijk\ell;\alpha}(c_{[p,q]}c_{[q,v]}c_{[v,p]})=\Delta_{ijk\ell;\alpha}(c_{[p,q]}c_{[q,v]}c_{[v,w]}c_{[w,p]})=0$.
Only two moves in the list remain to be investigated: $\{1,1,1,1\}~\to~\{1,0,1,0\}$ and
$\{1,0,1,0\}~\to~\{1,1,1,1\}$. These last two moves involve all indices in $S_{pqvw}$ but are both incompatible with the action of any edge swap.
\vsp

\item
For each $n=3$ term, there are three index pairs  $(x,y)\in\{[p,q],[q,v],[v,p]\}$ in the case of ${\rm Tr}(\bc^3)$ and three index pairs  $(x,y)\in\{[p,q],[q,v],[v,w],[w,p]\}$ in the case of ${\rm Tr}(\bc^4)$ such that $F_{ijkl;\alpha}c_{xy}=1-c_{xy}$. Since $F_{ijkl;\alpha}$ removes two bonds and adds two new ones, of these pairs $(x,y)$ at least one must have $c_{xy}=0$ and at least one must have $c_{xy}=1$. So the options for the action of $F_{ijkl;\alpha}$ are (modulo permutations):
\begin{eqnarray*}
{\rm Tr}(\bc^3):&~~&\{c_{[p,q]},c_{[q,v]},c_{[v,p]}\}=\{1,1,0\}~\to~\{0,0,1\}\\
                &~~&\{c_{[p,q]},c_{[q,v]},c_{[v,p]}\}=\{1,0,0\}~\to~\{0,1,1\}\\[2mm]
{\rm Tr}(\bc^4):&~~&\{c_{[p,q]},c_{[q,v]},c_{[v,w]},c_{[w,p]}\}=\{1,1,1,0\}~\to~\{1,0,0,1\}\\
&~~&\{c_{[p,q]},c_{[q,v]},c_{[v,w]},c_{[w,p]}\}=\{1,1,0,0\}~\to~\{1,0,1,1\}~{\rm or}~\{0,0,1,0\}\\
&~~&\{c_{[p,q]},c_{[q,v]},c_{[v,w]},c_{[w,p]}\}=\{1,0,0,0\}~\to~\{0,1,1,0\}
\end{eqnarray*}
In all cases one has at least one $c_{xy}=0$ before the swap and at least one $c_{xy}=0$ after the swap, hence all $n=3$ terms have
$\Delta_{ijk\ell;\alpha}(c_{[p,q]}c_{[q,v]}c_{[v,p]})=\Delta_{ijk\ell;\alpha}(c_{[p,q]}c_{[q,v]}c_{[v,w]}c_{[w,p]})=0$.
\vsp

\item
For each $n=4$ term, which occur only in ${\rm Tr}(\bc^4)$, all four index pairs $(x,y)\in\{[p,q],[q,v],[v,w],[w,p]\}$ refer to links mapped according to $F_{ijkl;\alpha}c_{xy}=1-c_{xy}$. Since $F_{ijkl;\alpha}$ removes two bonds and adds two new ones, we must have $
c_{[p,q]}+c_{[q,v]}+c_{[v,w]}+c_{[w,p]}=2$ both before and after the swap, so $c_{[p,q]}c_{[q,v]}c_{[v,w]}c_{[w,p]}=F_{ijk\ell;\alpha}(c_{[p,q]}c_{[q,v]}c_{[v,w]}c_{[w,p]})=0$ and
hence all $n=4$ terms have
$\Delta_{ijk\ell;\alpha}(c_{[p,q]}c_{[q,v]}c_{[v,w]}c_{[w,p]})=0$.
\vsp

\end{itemize}
It follows therefore that we may write, with the short-hands $A/B=\{x\in A|~x\notin B\}$ and $\overline{c}_{xy}=1-c_{xy}$,
\begin{eqnarray}
\Delta_{ijk\ell;\alpha}{\rm Tr}(\bc^3)&=& \sum_{pqv}
\delta_{|S_{pqv;ijk\ell;\alpha}|,1}
\Delta_{ijk\ell;\alpha}(c_{[p,q]}c_{[q,v]}c_{[v,p]})
\nonumber\\
&=&
\sum_{pqv}
\delta_{|S_{pqv;ijk\ell;\alpha}|,1}
\Big(\prod_{(x,y)\in S_{pqv}\!/\!S_{pqv;ijk\ell;\alpha}}\!c_{xy}\Big)
\Delta_{ijk\ell;\alpha}\Big(\prod_{(x,y)\in S_{pqv;ijk\ell;\alpha}}\!\!c_{xy}\Big)
\nonumber\\
&=&
\sum_{pqv}
\delta_{|S_{pqv;ijk\ell;\alpha}|,1}
\Big(\prod_{(x,y)\in S_{pqv}\!/\!S_{pqv;ijk\ell;\alpha}}\!\!c_{xy}\Big)
\Big(\prod_{(x,y)\in S_{pqv;ijk\ell;\alpha}}\!\overline{c}_{xy}-\!\!
\prod_{(x,y)\in S_{pqv;ijk\ell;\alpha}}\!c_{xy}\Big)~~~~~
\label{eq:3}
\end{eqnarray}
and
\begin{eqnarray}
\Delta_{ijk\ell;\alpha}{\rm Tr}(\bc^4)&=& \sum_{pqvw}
\delta_{|S_{pqvw;ijk\ell;\alpha}|,1}
\Delta_{ijk\ell;\alpha}(c_{[p,q]}c_{[q,v]}c_{[v,w]}c_{[w,p]})
\nonumber
\\
&=& \sum_{pqvw}
\delta_{|S_{pqvw;ijk\ell;\alpha}|,1}
\Big(\prod_{(x,y)\in S_{pqvw}\!/\!S_{pqvw;ijk\ell;\alpha}}\!c_{xy}\Big)
\Delta_{ijk\ell;\alpha}\Big(\prod_{(x,y)\in S_{pqvw;ijk\ell;\alpha}}\!\!c_{xy}\Big)
\nonumber\\
&=&
\sum_{pqvw}
\delta_{|S_{pqvw;ijk\ell;\alpha}|,1}
\Big(\prod_{(x,y)\in S_{pqvw}\!/\!S_{pqvw;ijk\ell;\alpha}}\!\!c_{xy}\Big)
\Big(\prod_{(x,y)\in S_{pqvw;ijk\ell;\alpha}}\!\overline{c}_{xy}-\!\!
\prod_{(x,y)\in S_{pqvw;ijk\ell;\alpha}}\!c_{xy}\Big)~~~~~
\nonumber
\\
{}
\label{eq:4}
\end{eqnarray}
The above expressions can be simplified once more, taking into account the number of ways we can select the common index pair $(a,b)$ from $S_{pqv}$ or $S_{pqvw}$, respectively, and the fact that each such pair will be picked up twice
(as $[a,b]$ and $[b,a]$, respectively) in the above summations, due to $c_{ab}=c_{ba}$:
\begin{eqnarray}
\Delta_{ijk\ell;\alpha}{\rm Tr}(\bc^3)&=&
2\cdot 3\sum_{[a,b]\in S_{ijk\ell;\alpha}}(1\!-\!2c_{ab})
\sum_{v\notin\{a,b,b^{[a,b]}_{+;\alpha},a^{[a,b]}_{-;\alpha}\}}
c_{bv}c_{va}=
6\sum_{[a,b]\in S_{ijk\ell;\alpha}}(1\!-\!2c_{ab})\sum_{v\notin\{i,j,k,\ell\}}
c_{bv}c_{va}\nonumber
\\
{}
\label{eq:Tr3term}
\\
\Delta_{ijk\ell;\alpha}{\rm Tr}(\bc^4)&=&
2\cdot 4
\sum_{[a,b]\in S_{ijk\ell;\alpha}}(1\!-\!2c_{ab})
\sum_{v\notin\{a,b,b^{[a,b]}_{+;\alpha}\}}
\left(
\sum_{w\notin\{a,b,a^{[a,b]}_{-;\alpha}\}, (v,w)\neq(a^{[a,b]}_{-;\alpha},
b^{[a,b]}_{+;\alpha})}
c_{bv}c_{vw}c_{wa}\right)
\label{eq:Tr4term}
\end{eqnarray}
The constraints over the sums in (\ref{eq:Tr3term}) and (\ref{eq:Tr4term})
implement the $\delta$s in (\ref{eq:3}) and (\ref{eq:4}), and guarantee
that no link other than $[a,b]$ can be picked up from $S_{ijk\ell;\alpha}$
(note that
$(b,a^{[a,b]}_{-;\alpha})$ and $(a,b^{[a,b]}_{+;\alpha})
\notin S_{ijk\ell;\alpha} ~\forall\alpha$).

We can verify briefly our results (\ref{eq:Tr3term},\ref{eq:Tr4term}) for the nearly hardcore graphs, where the
variation in the mobility term can be calculated explicitly.
Referring back to Figure \ref{fig:softcore},
the difference between the mobilities of a B type graph and graph A is
\be
\Delta_{ijk\ell;\alpha}n({\bf c})=n_B({\bf c})-n_A({\bf c})=
K^2-K-2K+2=K^2-3K+2
\ee
If we identify sites $K+1=j$ and $K+2=i$,
we see that graph A is obtained from graph B by application
of the edge swap $F_{ijk\ell;1}$.
Hence, the variation in the mobility produced by the application of
$F_{ijk\ell;1}$ to graph $B$ must be equal, via (\ref{eq:nc}), to
\be
\Delta_{ijk\ell;1}n({\bf c})=
\frac{1}{4}\Delta_{ijk\ell;1}{\rm Tr}(\bc^4)+
\frac{1}{2}\Delta_{ijk\ell;1}{\rm Tr}(\bc^3)
-\frac{1}{2}\sum_{vw}k_vk_w\Delta_{ijk\ell;1}c_{vw}
\ee
with $\Delta_{ijk\ell;1}{\rm Tr}(\bc^4)$ and
$\Delta_{ijk\ell;1}{\rm Tr}(\bc^3)$ given by (\ref{eq:Tr3term}) and
(\ref{eq:Tr4term})
for $\alpha=1$.

For the nearly hardcore case one can see that the only link
$[a,b]\in S_{ijk\ell;1}$ contributing to the sums in (\ref{eq:Tr3term})
and (\ref{eq:Tr4term}) is $[k,\ell]$,
because if $v\notin\{a,b^{[a,b]}_{+;1}\}$ then $c_{bv}=0$ for $b\in\{i,j\}$,
and if $w\neq b$ then $c_{wa}=0$ for $a=j$;
so the links $[a,b]\in\{[i,j],[j,k],[\ell,i]\}$ never contribute. Using
$c_{kl}=0$,
we have
\bea
\Delta_{ijk\ell;1}n({\bf c})&=&2\sum_{v\notin \{k,l,i\}}
\left(\sum_{w\notin\{k,l,j\}; (v,w)\neq(j,i)}c_{\ell v}c_{vw}c_{wk}\right)
+3 \sum_{v\notin \{ijk\ell\}}c_{\ell v}c_{vk}
-\frac{1}{2} \sum_{\{v,w\}=\{i,j\};v\neq w}\Delta_{ijk\ell;1}c_{vw}
\nonumber\\
&&-2\cdot \frac{1}{2}(K-1)
\sum_{v\in\{i,j\}}\sum_{w=1}^K\Delta_{ijk\ell;1}c_{vw}
-\frac{1}{2}(K-1)^2\left(\sum_{\{v,w\}\neq\{k,\ell\};v\neq w}^K
\Delta_{ijk\ell;1}c_{vw}\right)
\nonumber\\
&=& 2(K-2)(K-3)+3(K-2)-1+2(K-1)-(K-1)^2=K^2-3K+2
\eea
where in the last equality we used the fact that
$\Delta_{ijk\ell;1}c_{vw}=1-2c_{vw}$ if $(v,w)\in S_{ijkl}$,
with $\Delta_{ijk\ell;1}c_{vw}=0$ otherwise, as well as  $c_{jk}=c_{\ell i}=1$ and $c_{ij}=0$.

\subsection{Bounds for $\Delta_{ijk\ell;\alpha}{\rm Tr}(\bc^3)$ and $\Delta_{ijk\ell;\alpha}{\rm Tr}(\bc^4)$}

It is easy to construct  bounds from (\ref{eq:Tr3term},\ref{eq:Tr4term}), based on the property $1-2c_{ab}=\pm 1$, on the fact that always precisely two of the four bonds in $S_{ijk\ell;\alpha}$ are zero, and on the inequality $xy\leq \frac{1}{2}(x+y)$ for $x,y\in\{0,1\}$:
 \begin{eqnarray}
|\Delta_{ijk\ell;\alpha}{\rm Tr}(\bc^3)|&\leq &
12\max_{a\neq b}\sum_v c_{av}c_{vb} \leq 6\max_{a\neq b}\sum_v(c_{av}+c_{vb})
\leq 12 k_{\rm max}
\label{eq:Tr3_bound}
\\
|\Delta_{ijk\ell;\alpha}{\rm Tr}(\bc^4)|&\leq &
16 \max_{a\neq b}\sum_{v,w\notin\{a,b\}} c_{av}c_{vw}c_{wb}\leq 8 \max_{a\neq b}\sum_{v,w\notin\{a,b\}}c_{vw}(c_{av}+c_{wb})
\nonumber
\\
&\leq& 16 \max_{i}\Big(\sum_{j}c_{ij}k_{j}\Big)
\label{eq:Tr4_bound}
\end{eqnarray}
with $k_{\rm max}=\max_i k_{i}$.
We can finally simplify (\ref{eq:Tr4_bound}) in two ways. First, we may use $k_j\leq \max_j k_j$, which gives $\max_i [\sum_j c_{ij} k_j]\leq k_{\rm max}^2$. Second, we could simply put $\max_i [\sum_j c_{ij} k_j]\leq [\sum_j k_j]=N\bra k\ket$. The result is:
\begin{eqnarray}
|\Delta_{ijk\ell;\alpha}{\rm Tr}(\bc^4)|&\leq & 16 \min\{k_{\rm max}^2,N\bra k\ket\}
\label{eq:Tr4_bound2}
\end{eqnarray}

\end{document}